\documentclass[aps,twocolumn,superscriptaddress]{revtex4-1}

\usepackage{breqn}
\usepackage{graphicx}
\usepackage{dcolumn}   
\usepackage{bm}        
\usepackage{amssymb}   
\usepackage{soul}
\usepackage{gensymb}
\usepackage{enumerate}
\usepackage{multirow}
\usepackage[T1]{fontenc}

\usepackage[breaklinks=true,colorlinks=true,linkcolor=blue,urlcolor=blue,citecolor=blue]{hyperref}

\usepackage{float}

\begin{document}

\title{First global gyrokinetic profile predictions of ITER burning plasma}
\author{A.~Di Siena} 
\affiliation{Max Planck Institute for Plasma Physics Boltzmannstr 2 85748 Garching Germany}
\author{C.~Bourdelle} 
\affiliation{CEA IRFM F13108 Saint Paul lez Durance France}
\author{A.~Ba\~n\'on~Navarro}
\affiliation{Max Planck Institute for Plasma Physics Boltzmannstr 2 85748 Garching Germany}
\author{G.~Merlo} 
\affiliation{Max Planck Institute for Plasma Physics Boltzmannstr 2 85748 Garching Germany}
\author{T.~G\"orler}
\affiliation{Max Planck Institute for Plasma Physics Boltzmannstr 2 85748 Garching Germany}
\author{E.~Fransson}
\affiliation{CNRS AixMarseille Univ PIIM UMR7345 Marseille France}
\author{A.~Polevoi}
\affiliation{ITER Organization Route de Vinon sur Verdon  CS 90 046 13067 St Paul lez Durance Cedex France}
\author{S.~H.~Kim}
\affiliation{ITER Organization Route de Vinon sur Verdon  CS 90 046 13067 St Paul lez Durance Cedex France}
\author{F.~Koechl}
\affiliation{ITER Organization Route de Vinon sur Verdon  CS 90 046 13067 St Paul lez Durance Cedex France}
\author{A.~Loarte}
\affiliation{ITER Organization Route de Vinon sur Verdon  CS 90 046 13067 St Paul lez Durance Cedex France}
\author{E.~Fable}
\affiliation{Max Planck Institute for Plasma Physics Boltzmannstr 2 85748 Garching Germany}
\author{C.~Angioni}
\affiliation{Max Planck Institute for Plasma Physics Boltzmannstr 2 85748 Garching Germany}
\author{P.~Mantica}
\affiliation{Istituto per la Scienza e Tecnologia Dei Plasmi Consiglio Nazionale Delle Ricerche Milano Italy}
\author{F.~Jenko} 
\affiliation{Max Planck Institute for Plasma Physics Boltzmannstr 2 85748 Garching Germany}

\begin{abstract}

In this work, we present the first global gyrokinetic simulations of the ITER baseline scenario operating at 15MA using GENE-Tango electrostatic and electromagnetic simulations. The modeled radial region spans close to the magnetic axis up to $\rho_{tor}=0.6$.  The fusion power is self-consistently evolved, while the interplay of alpha particles with turbulence left for future work. Our results show a pronounced density peaking, moderated by electromagnetic fluctuations. The predicted fusion gain for this scenario is $Q=12.2$, aligning well with ITER’s mission objectives. We further characterize the turbulence spectra and find that electromagnetic modes, such as microtearing modes (MTMs), kinetic ballooning modes (KBMs) and Alfv\'enic ion temperature gradient modes (AITGs) at low bi-normal wave numbers, play a critical role in the core transport of this ITER scenario, necessitating high numerical resolution for accurate modeling. Local flux-tube simulations qualitatively reproduce the key features observed in the global gyrokinetic simulations but exhibit a much higher sensitivity to profile gradients—reflecting increased stiffness, likely due to the linearization of the equilibrium profiles and safety factor, which strongly influence the drive of these electromagnetic modes. Our study also reveals that the imposed external toroidal rotation profiles have a negligible impact on turbulent transport, as their magnitudes are substantially lower than the dominant linear growth rates. Furthermore, we demonstrate that the safety factor profile is of paramount importance: scenarios featuring flat $q$ profiles with near-zero magnetic shear lead to the destabilization of kinetic ballooning modes (KBMs) in the plasma core, significantly enhancing turbulent transport and potentially degrading confinement. Finally, although electron temperature gradient (ETG) turbulence initially appears large, sometimes exceeding ion-scale transport levels, it is ultimately quenched over long timescales by secular evolution of zonal flows, which are weakly damped under the very low collisionality conditions expected in ITER.

\end{abstract}


\maketitle


\section{Introduction}

The construction of ITER in southern France continues to advance, with a revised plan to start DT operations following the Start of Research Operation (SRO) and post-SRO assembly phases in 2030s. ITER is designed to achieve burning plasma conditions where the energy produced by fusion reactions is at least ten times greater than the external power input \cite{Loarte_NF_2025}, meaning alpha heating twice larger than auxillary power. This milestone would mark a pivotal achievement in establishing fusion as a viable energy source, making reliable predictions for ITER’s performance critically important. Such predictive studies enable exploration of plasma behavior under a range of conditions before actual experiments are performed, helping to anticipate and mitigate potential confinement degradation that could compromise ITER’s objectives. The ITER Research Plan outlines a variety of these operational scenarios. Among them is the baseline scenario, characterized by a plasma current of 15MA and an external heating power of 50MW \cite{Kim_NF_2018,Mantica_PPCF_2020, Citrin_PoP_2023,Eriksson_2024, MilitelloAsp_2022}. Additionally, ITER is expected to explore advanced scenarios \cite{Gormezano_NF_2007,Shimada_NF_2007,Kim_NF_2016,Bock_NF_2017,Polevoi_NF_2020,Kim_NF_2021,Kim_NF_2024,Reisner_NF_2025} as well, aimed at achieving lower fusion power with $Q = 5$ but prolonged burning phases. These typically operate at lower plasma currents, feature a higher fraction of non-inductive current drive, and pursue improved confinement alongside elevated plasma beta ($\beta$, defined as the ratio of kinetic to magnetic pressure) to advance the understanding and optimization of fusion performance.

In recent years, considerable effort has been spent performing such predictive simulations, mostly using reduced quasilinear turbulence models that offer immense computational speedups compared to high-fidelity gyrokinetic simulations, the two main models are QuaLiKiz and TGLF, see the recent review \cite{Staebler_2024} and references therein. The quasilinear models are embedded in integrated modelling frameworks such as ASTRA, HFPS, etc, see \cite{Bourdelle_NF_2005} for a review of such frameworks and recent validation results. 

However, the degree of prediction that can be achieved in the ITER burning plasma simulations remains rather uncertain with quasilinear turbulence models. This concern is underscored by the fact that even in present-day devices, there are parameter ranges, such as high-beta plasmas or plasmas with a significant population of supra-thermal particles, where these reduced models fail to accurately reproduce experimental observations \cite{Doerk_NF_2017, Casson_2020, Citrin_PPCF_2023, DiSiena_NF_2024}. These limitations highlight the need for further investigations using higher-fidelity tools to predict plasma behavior in future fusion devices like ITER, as the saturation rule, developed for present-day devices, needs to be verified for next-generation machines. While such simulations were previously prohibitively expensive, recent advances in GPU porting (see e.g., Refs.~\cite{Germaschewski_PoP_2021}, and~\cite{Mandell_JPP_2024}) and HPC clusters have dramatically lowered the computational cost, enabling, for the first time, global electromagnetic gyrokinetic simulations of burning plasmas relevant to ITER.

This led to the development of sophisticated modeling frameworks that couple gyrokinetic turbulence simulations with transport solvers. Examples include GS2-Trinity, GENE-Trinity \cite{Barnes_PoP_2010}, and CGYRO combined with nonlinear optimization techniques based on Gaussian process regression \cite{Rodriguez_Fernandez_2022,Fernandez_NF_2024,Howard_NF_2025}. Beyond local flux-tube simulations, efforts have also focused on coupling global gyrokinetic models, such as the global version of GENE \cite{Jenko_PoP2000,Goerler_JCP2011}, to transport solvers like Tango \cite{Shestakov_JCP_2003, Parker_NF_2018, DiSiena_NF_2022, Disiena_NF}. This approach enables the inclusion of crucial radially global phenomena that influence turbulent transport, including profile shearing \cite{Garbet_PoP_1996,Waltz_PoP_2005,Hornsby_NF_2018}, avalanche-like transport events \cite{Candy_PRL_2003,Sarazin_PoP_2000}, the emergence of internal transport barriers \cite{Strugarek_PPCF_2013,DiSiena_PRL_2021,Di_Siena_PPCF_2022}, turbulence spreading \cite{Hahm_PPCF_2004}, and the role of fusion born alpha particles \cite{Chen_RMP_2016, DiSiena_NF_2023_2}. The GENE-Tango framework has previously been validated against experimental data from multiple ASDEX Upgrade discharges. These include cases with moderate external heating power, where the influence of energetic particles on core profiles was negligible \cite{DiSiena_NF_2022}, as well as high-power discharges in which energetic particle effects were found to be significant \cite{DiSiena_NF_2024}. Interestingly, in such regimes, GENE-Tango successfully reproduced the experimentally observed ion temperature peaking, whereas reduced turbulence models such as TGLF-ASTRA predicted much flatter profiles. In addition, GENE-Tango has been validated at JET in D-T plasmas, successfully modeling the discharge with the highest fusion performance across both the DT1 and DT2 experimental campaigns in 50-50 D-T plasmas, with excellent agreement between simulations and measurements \cite{DiSiena_NF_2025}. More recently, the GENE-Tango framework has also been extended to optimized stellarator configurations \cite{Navarro_NF_2023} and validated across multiple OP1.2b W7-X scenarios \cite{Fernando_inpreparation}.

In this work, we exploit the coupling between the global version of the gyrokinetic code GENE and the transport solver Tango to carry out, for the first time, first-principles, radially global gyrokinetic simulations of the ITER baseline scenario. Through this approach, we predict the stationary plasma profiles for this scenario in fully electromagnetic simulations. In particular, we assess the impact of finite-$\beta$ effects, the role of electromagnetic modes, toroidal plasma rotation, the safety factor profile, and ETG turbulence on the overall plasma performance of this ITER baseline scenario.

This paper is organized as follows. Section \ref{sec1} provides a detailed description of the ITER baseline scenario analyzed in this work. Section \ref{sec2} outlines the numerical setup used for the GENE-Tango simulations, while Section \ref{sec3} briefly discuss the coupling between the global gyrokinetic code GENE and the transport solver Tango. In Section \ref{sec4}, we present steady-state plasma profiles obtained from various GENE-Tango simulations in the electrostatic and electromagnetic limits. Section \ref{sec5} offers a detailed analysis of the plasma instabilities present in these steady-state profiles, with comparisons to local flux-tube simulations discussed in Section \ref{sec6}. Section \ref{sec7} presents a comparison with plasma profiles calculated using reduced turbulence models. The role of plasma rotation on turbulent fluxes in radially global GENE simulations is investigated in Section \ref{sec8}, followed by an assessment of the impact of a nearly flat safety factor profile in the plasma core in Section \ref{sec9}. Additionally, Section \ref{sec10} evaluates the influence of ETG turbulence on the final steady-state plasma profiles. Finally, conclusions are drawn in Section \ref{sec11}.

\section{ITER 15MA scenario description} \label{sec1}

Among the various operational regimes envisioned for ITER, one of the most important is the so-called ITER baseline scenario designed to controll burning plasma conditions at high fusion power and power multiplication factor. This scenario targets a fusion power output of at least 500MW, corresponding to a fusion gain factor of $Q = 10$, under conditions of 15MA plasma current and a safety factor $q_{95} \approx 3$. This plasma scenario has been widely explored in the literature to estimate plasma performance and optimize profile evolution \cite{Citrin_NF_2010,Mantica_PPCF_2019, Koechl_NF_2020, MilitelloAsp_2022, Angioni_NF_2023,Citrin_PoP_2023,Eriksson_2024, Luda_NF_2025,Howard_NF_2025}. In this study, we extend these efforts by employing the GENE-Tango framework to perform physics-based, first-principles simulations of this scenario using global gyrokinetic simulations for the first time. Our approach leverages global electromagnetic gyrokinetic simulations coupled with a transport solver to predict the self-consistent evolution of temperature and density profiles from $\rho_{tor}=0.6$ inward. The initial and boundary conditions as well as the non-evolving quantities (the plasma composition, the toroidal rotation, the NBI and ECRH sources) have been prepared using the modelled ITER 15 MA plasma presented on Fig.~8 of \cite{Mantica_PPCF_2020} and further detailed in \cite{koechl2018}. The simulation was carried out using the HFPS integrated modeling framework, the reduced turbulent transport model used to predict heat and particle fluxes is QuaLiKiz, the ECRH power (20 MW) is a Gaussian centered at $\rho_{tor}=0.4$, the NBI source (33 MW) is self-consistently calculated by PENCIL \cite{Challis_NF_1989}. The density at the pedestal top is fixed and the pellet is modeled using a Gaussian centered at $\rho_{tor}=0.85$, the temperature at the pedestal top is adjusted so that the pedestal pressure agrees with the EPED based expectation \cite{Polevoi_2015} of 130 kPa. The toroidal rotation profile, kept fixed in the present work, has been obtained by using QuaLiKiz with a Prandtl number of 1. The effective charge results from He, Ne, Be and W, transported by QuaLiKiz and subsequently kept fixed (around 1.5 see Fig.~\ref{fig:fig_zeff}) in the GENE-Tango framework where it impacts only the collisionality. The above described HFPS-QuaLiKiz simulation has been chosen as reference in this work because in \cite{Citrin_PoP_2023} the QuaLiKiz predicted temperature and density profiles within $\rho=0.92$ were shown to agree well with a set of neural
networks trained on a quasilinear electrostatic GENE dataset, with a saturation rule calibrated to dedicated nonlinear simulations (Fig.~12 of \cite{Citrin_PoP_2023}). The only main modification compared with the initial HFPS-QuaLiKiz simulation is that we used a relaxed $q$ profile rather than the flat one inside $q=1$ from \cite{koechl2018}. The impact of the safety factor on the electromagnetic turbulent flux will be discussed in section \ref{sec8}. Note that the impact of the fixed toroidal rotation is also discussed later in section \ref{sec7}.

\section{Numerical setup and grid resolution} \label{sec2}

The turbulence simulations presented in this study are carried out using the GENE code \cite{Jenko_PoP2000, Goerler_JCP2011}, a Eulerian gyrokinetic code that evolves the Vlasov-Maxwell system of equations on a five-dimensional grid in field-aligned coordinates. The spatial coordinates $(x,y,z)$ correspond respectively to the radial, bi-normal at outboard midplane, and field-aligned directions, while the velocity space is represented by $(v_\shortparallel,\mu)$, denoting the velocity component parallel to the background magnetic field and the magnetic moment. GENE uses a $\delta$f approach, splitting the distribution function into a static equilibrium part and a time-dependent fluctuating component, and solves the equations for the latter. GENE includes electromagnetic fluctuations (both parallel and perpendicular magnetic field components \cite{Sheffield_PPCF_2025}), realistic magnetic geometries, multiple species, different options for modeling collisions, external $E\times B$ shear, and various MHD models.

GENE supports both flux-tube (local) and radially global gyrokinetic simulations. Flux-tube gyrokinetic simulations are based on the assumption that turbulent transport is predominantly local in nature and can be accurately captured by solving the nonlinear Vlasov–Maxwell system within a narrow annulus of the plasma centered around a specific flux surface. Within this domain, background profiles are treated as constant, while retaining their gradients. The flux-tube version of the code assumes periodic boundary conditions in both radial and bi-normal directions, enabling a Fourier representation in these coordinates. In contrast, global gyrokinetic simulations cover a broader radial domain and keep the full radial variation of equilibrium quantities. As a result, periodic boundary conditions can no longer be applied in the radial direction, and a Fourier decomposition cannot be used. To sustain the background profiles throughout the simulation, global codes either employ artificial Krook-type sources in gradient-driven setups or implement physical sources in flux-driven simulations.

The flux-tube version of the code is generally numerically less demanding than the global one and is often used to simulate plasma conditions with much lower computational cost, while maintaining a level of accuracy comparable to that of the more expensive global simulations. However, certain scenarios necessitate the use of the global gyrokinetic simulations, such as cases involving turbulence avalanches, strong electromagnetic turbulence, or energetic particle-driven modes. In these situations, the assumptions of the local approximation can fail to capture important dynamics, leading to significant differences between local and global results (see, e.g., Ref.~\cite{McMillan_PRL_2010,DiSiena_NF_2023_2}). Unless otherwise stated, the simulations presented in this work are performed using the global version of the GENE code.

All simulations presented in this study use a realistic ion-to-electron mass ratio and include electromagnetic effects (only the $A_\shortparallel$ component). Collisions are modeled using a linearized Landau-Boltzmann operator that conserves both energy and momentum \cite{Crandall_CPC_2020} unless stated otherwise. Additionally, external toroidal rotation is also considered (see Fig.~\ref{fig:fig_rotation}.) and the magnetic geometry is computed using CHEASE \cite{Lutjens_CPC_1996} to be self-consistent with the plasma profiles (see discussion in Sec.~\ref{sec3}). The simulations are carried out over the radial interval $\rho_{\text{tor}} = [0.075, 0.6]$, where $\rho_{\text{tor}} = \sqrt{\Phi / \Phi_{\text{LCFS}}}$ represents the normalized toroidal flux. Unless stated otherwise, the spatial grid resolution is set to $(n_x \times n_{k_y} \times n_z) = (1024 \times 64 \times 32)$, corresponding to the radial $(x)$, bi-normal $(y)$, and field-aligned $(z)$ directions, respectively. The toroidal mode numbers are discretized using the relation $n = n_{0,\text{min}} \cdot j$, with $j$ taking integer values in the range $0 \leq j \leq n_{k_y}-1$, and $n_{0,\text{min}} = 6$. In normalized units, the minimum toroidal mode number at the center of the radial GENE domain corresponds to $k_y \rho_s = 0.022$ while the maximum to $k_y \rho_s \approx 1.4$. This relatively low value is necessary to accurately capture low-$k_y$ electromagnetic modes, which play a non-negligible contribution to the turbulent transport. It is worth mentioning that while capturing the $n = 1$ mode could be relevant, especially given that the safety factor $q$ reaches unity, doing so would require a sixfold increase in bi-normal resolution, making the simulations computationally prohibitive within the available resources. The velocity grid spans $l_v = [-3.5, 3.5]$ thermal velocities along the direction parallel to the magnetic field and $l_\mu = [0, 12.5]$ in the magnetic moment, normalized to GENE units at the radial center of the domain with a resolution, respectively, of $(n_{v_\shortparallel} \times n_{\mu}) = (48 \times 32)$.

All simulations presented in this manuscript are carried out in the gradient-driven setup, where Krook-type operators for both particles and heat are applied to maintain the plasma profiles close to their initial values. 
The heat Krook coefficients are nearly an order of magnitude smaller than the maximum growth rate \cite{McMillan_PRL_2010}. To satisfy Dirichlet boundary conditions, buffer regions spanning $10\%$ of the radial domain are implemented near the simulated boundaries. Within these regions, a Krook damping operator with $\gamma_b = 1.0 \cdot c_s/a$ is applied to suppress fluctuations. Furthermore, a fourth-order hyperdiffusion scheme is employed along the radial and bi-normal direction \cite{Pueschel_CPC_2010} to damp unresolved fine electron scale turbulence due to electron temperature gradient modes.

To reduce the computational cost of the GENE simulations, the thermal deuterium and tritium species (assumed as a 50-50 mixture) are modeled as a single effective mixed-ion species, rather than being treated separately. This simplification has no impact on the linear results and is assumed to be valid also in the nonlinear regime, consistent with previous nonlinear GENE studies investigating the role of alpha particles at JET \cite{DiSiena_2025}. Impurities and fusion-born alpha particles are not included in the simulations. Assessing their influence on plasma performance and fusion power output is left for future work.

\section{Coupled GENE-Tango Simulations for Predictive Transport in the ITER Baseline Scenario} \label{sec3}

In this work, we perform plasma profile predictions for the ITER baseline scenario using the gyrokinetic code GENE and the transport solver Tango, via the integrated GENE-Tango framework \cite{Shestakov_JCP_2003,Parker_NF_2018,DiSiena_NF_2022}. GENE-Tango allows for accurate and efficient profile evolution by significantly reducing the computational time compared to conventional flux-driven gyrokinetic simulations, which typically require a simulation duration of the order of several energy confinement times \cite{DiSiena_NF_2022}.

\begin{figure*}
\begin{center}
\includegraphics[scale=0.35]{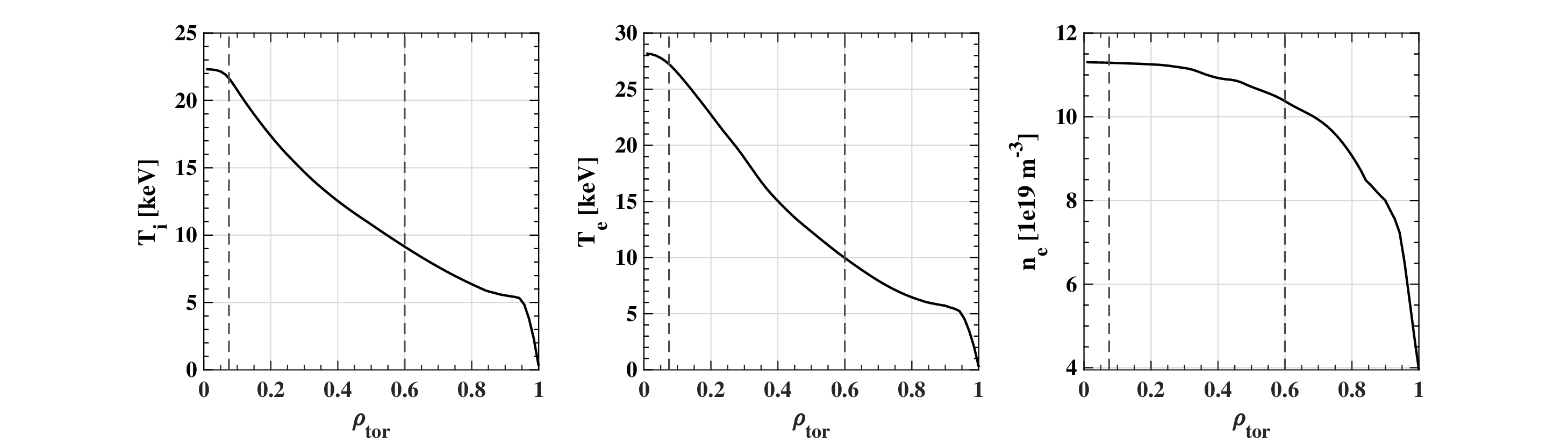}
\par\end{center}
\caption{Radial profiles of the initial (a) thermal ion temperature (50-50 deuterium-tritium mixture), (b) electron temperature, and (c) density for the ITER 15MA baseline scenario, as computed by QuaLiKiz-HFPS and used as initial conditions for the GENE-Tango simulations. The vertical dotted black lines denote the radial domain covered by the GENE-Tango simulations.}
\label{fig:fig_prof}
\end{figure*}

\begin{figure*}
\begin{center}
\includegraphics[scale=0.35]{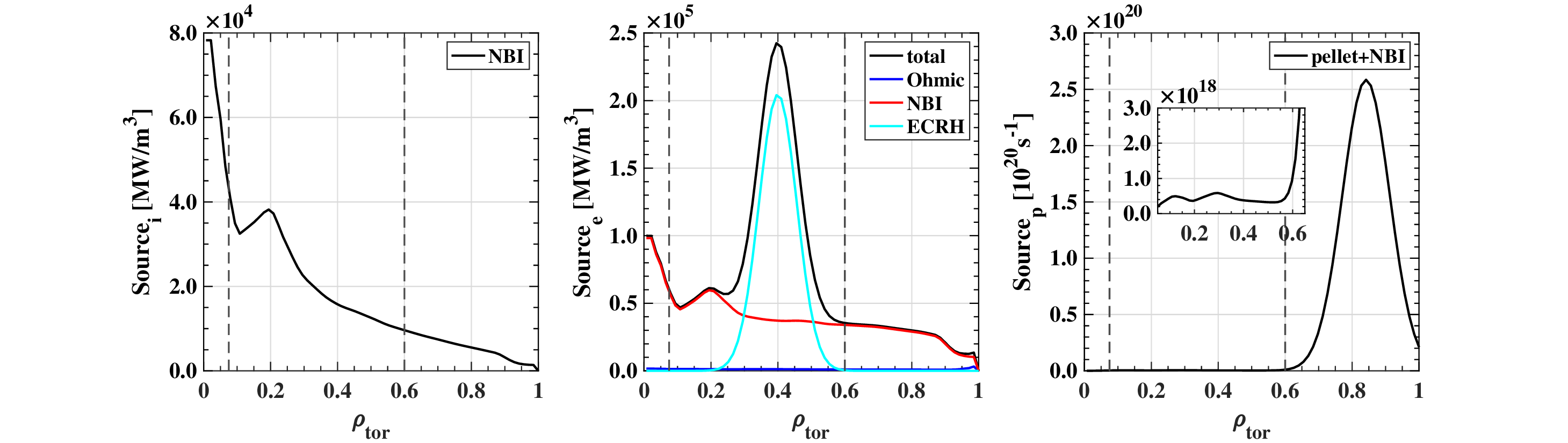}
\par\end{center}
\caption{Radial profiles of (a) ion heating, (b) electron heating, and (c) particle sources for the ITER 15MA baseline scenario, as computed by QuaLiKiz-HFPS and held fixed throughout the GENE-Tango simulations. Alpha particle heating, collisional energy exchange, and radiative power losses are computed self-consistently by GENE-Tango at each iteration.}
\label{fig:fig_sources}
\end{figure*}

\begin{figure*}
\begin{center}
\includegraphics[scale=0.35]{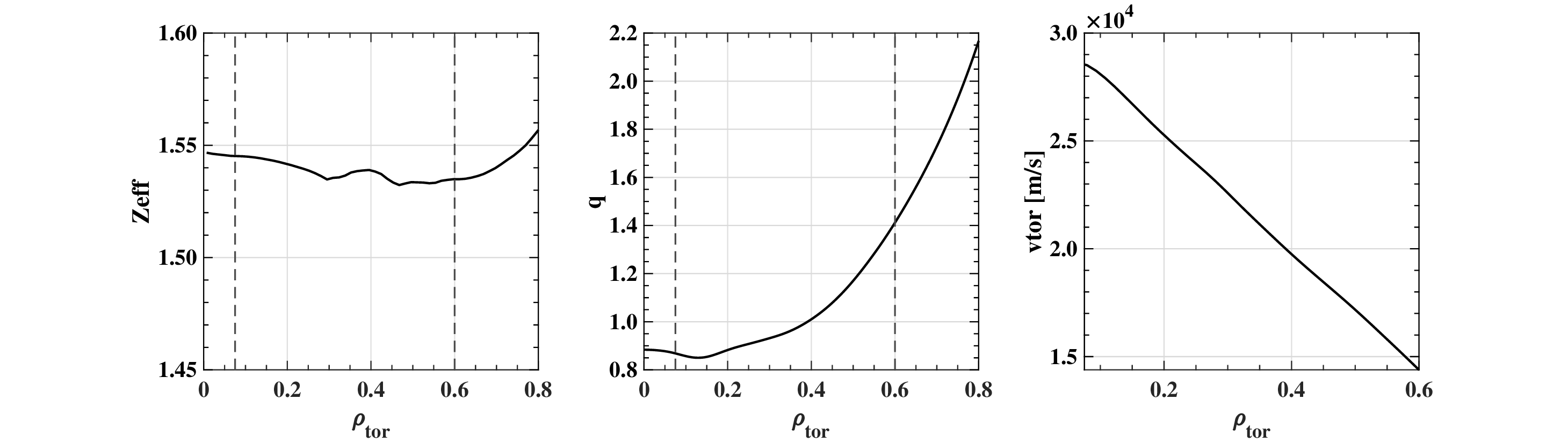}
\par\end{center}
\caption{Radial profiles of (a) $Zeff$, (b) safety factor $q$ and (c) toroidal rotation $vtor$ for the ITER 15MA baseline scenario, as computed by QuaLiKiz-HFPS and kept fixed throughout the GENE-Tango simulations.}
\label{fig:fig_zeff}
\end{figure*}

All the GENE-Tango coupled simulations presented in this work start with a standalone global GENE run, initialized using temperature and density profiles for each species taken from a prior HFPS-QuaLiKiz simulation detailed in \ref{sec1}. These input profiles are shown in Fig.~\ref{fig:fig_prof}. The rotation profile, the $q$ profile and the $Z_{\rm eff}$ profile are kept fixed and shown respectively in Fig.~\ref{fig:fig_rotation} (blue line) and Fig.~\ref{fig:fig_zeff}.
The GENE standalone simulations are run until the system reaches a saturated turbulent state. The resulting turbulent fluxes, averaged over the saturation phase, are then transferred to Tango, which evolves the plasma profiles (temperature and density for each species) self-consistently with the turbulent fluxes and the heat and particle sources. For a detailed description see Ref.~\cite{DiSiena_NF_2022}. The external sources used in Tango include Ohmic heating, electron cyclotron resonance heating (ECRH), and neutral beam injection (NBI) heating, as well as NBI fueling and pellets for the particle channels. These sources are kept fixed to the HFPS-QuaLiKiz simulation (see Sec.~\ref{sec1}) throughout the GENE-Tango iterations. No major change in the sources is expected during the profile evolution, assuming that the profiles do not undergo drastic variations. Their radial profiles are shown in Fig.~\ref{fig:fig_sources}.
In addition, Tango self-consistently calculates the collisional energy exchange, alpha heating, and radiation losses, including bremsstrahlung, line radiation, and synchrotron emission, based on the updated plasma profiles. For the calculation of these quantities, Tango takes into account all impurity species included in the HFPS-QuaLiKiz reference simulation, namely helium ash, beryllium, tungsten, neon, and alpha particles, resulting in an effective Zeff profile, as shown in Fig.~\ref{fig:fig_zeff}a.
On the other hand, GENE-Tango does not transport the impurity species in the simulations to reduce the computational cost of these simulations. Instead, it retains the $Z_{\rm eff}$ profile, which affects the collisional frequencies.

\begin{figure}
\begin{center}
\includegraphics[scale=0.40]{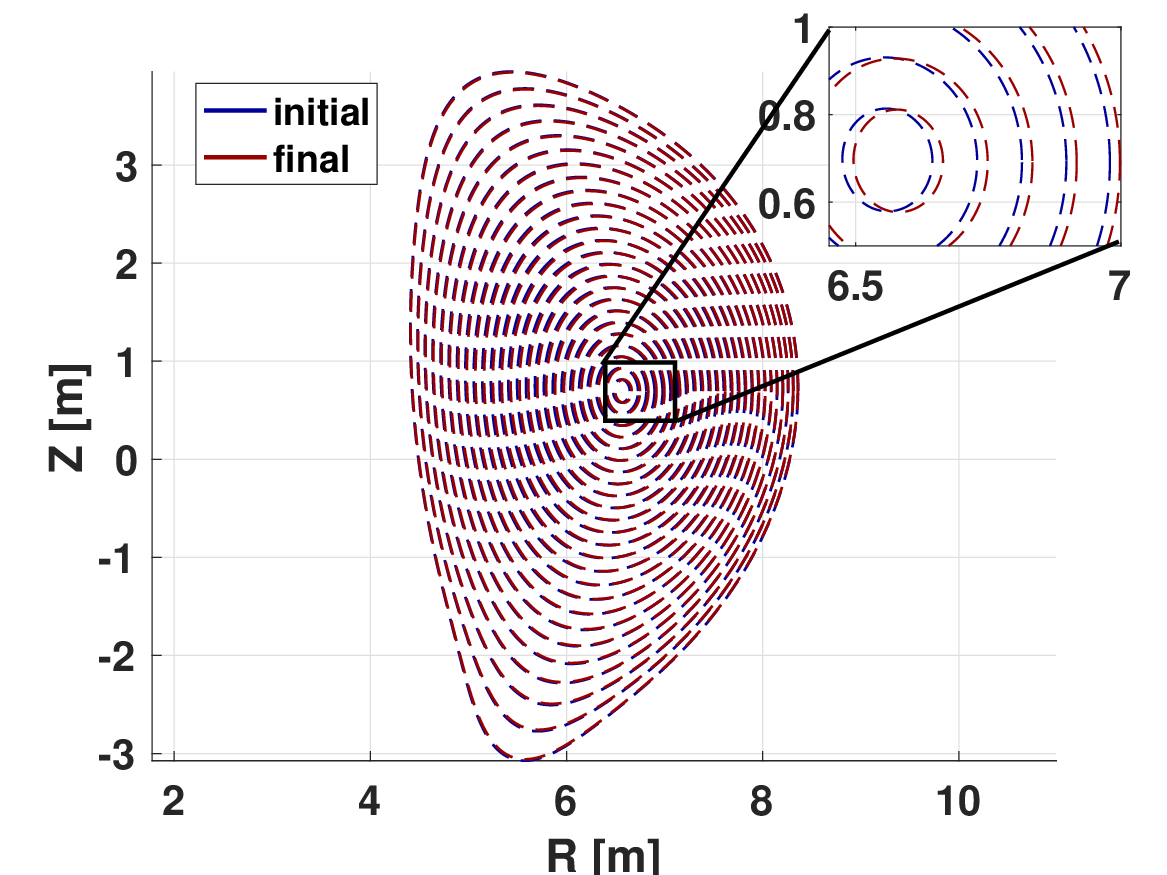}
\par\end{center}
\caption{Contours of constant poloidal flux from magnetic equilibria computed using CHEASE, based on the initial profiles from QuaLiKiz-HFPS (blue) and the final profiles from the GENE-Tango electromagnetic simulation (red).}
\label{fig:fig_RZ}
\end{figure}

To account for the changes in magnetic geometry resulting from the variations in plasma pressure, the geometry is recalculated every five GENE-Tango iterations using the updated pressure profile. This is done while keeping the safety factor profile fixed, based on the assumption that current diffusion occurs on longer timescales. The safety factor profile used is shown in Fig.~\ref{fig:fig_zeff}b. The updated magnetic equilibrium is computed using CHEASE, which runs with the fixed last closed flux surface from the initial equilibrium. The resulting geometry, along with the updated plasma profiles, are then transferred back to GENE. The simulation restarts from the previously reached saturated turbulent state and continues until a new saturated phase is achieved. A comparison between the flux surfaces of the initial and final magnetic geometries is shown in Fig.~\ref{fig:fig_RZ}.
The results indicate that the modifications in the plasma profiles have a limited impact on the magnetic geometry, leading to an outward shift of the magnetic axis by approximately 2 cm (a similar value is recovered when comparing the initial geometry to the electrostatic simulation). In the present study, each iteration is run for $t=150 c_s/a$, which is found to be sufficient to reach a new saturated turbulent state. The turbulent fluxes are then averaged over the later time window $t=[30-150] c_s/a$ within the $t=150 c_s/a$ to minimize the impact of the transient phase resulting from the change in the plasma profiles. This iterative process continues until the turbulent fluxes match the volume-averaged particle and energy sources. 

The global GENE simulations were performed on the Leonardo GPU cluster, using NVIDIA Ampere A100 GPUs. These simulations required 60,000 node-hours. This estimate accounts only for the main GENE-Tango production runs and does not include the additional computational time spent on numerical tests to determine optimal resolution or on further analyses performed on the simulation data.

\section{Electrostatic and Electromagnetic Transport Modeling with GENE-Tango} \label{sec4}

The numerical scheme described in the previous sections is applied here to predict plasma performance for the ITER baseline scenario and to assess the impact of electromagnetic effects on plasma confinement and fusion output. This is achieved by performing two GENE-Tango simulations: one in the electrostatic limit and the other including finite-beta electromagnetic effects. 
\begin{figure*}
\begin{center}
\includegraphics[scale=0.35]{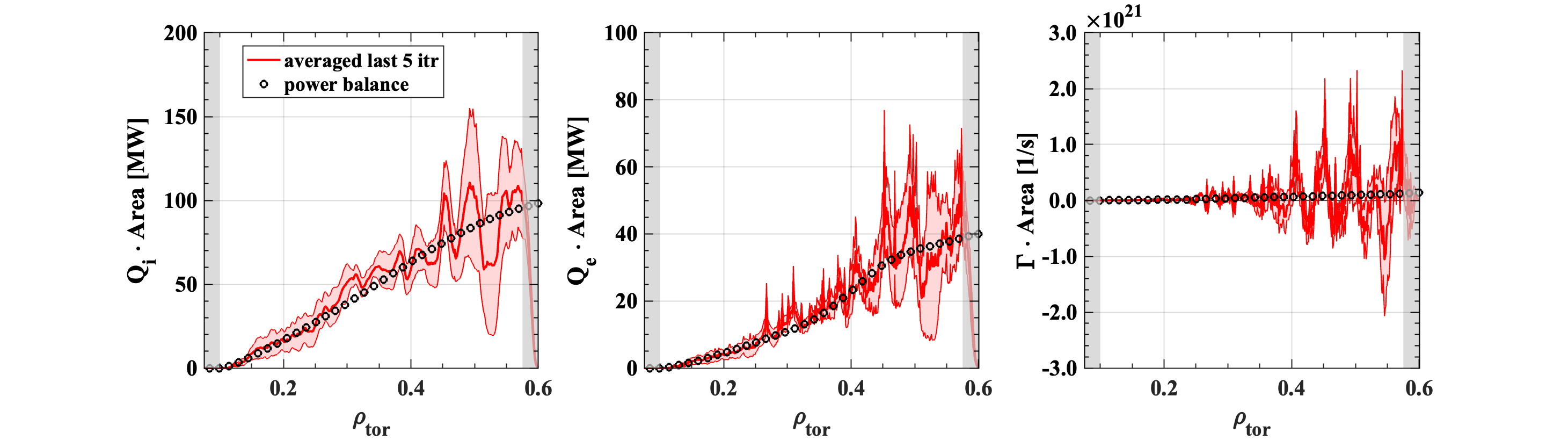}
\par\end{center}
\caption{Time-averaged radial profile of the (a) ion, (b) electron heat fluxes in MW and (c) particle flux in $1/s$ corresponding to the averaged last five GENE–Tango iterations (red) in the electrostatic GENE-Tango simulation. The shaded gray areas denote the buffer regions and the black circles the volume integral of the injected particle and heat sources. The shaded red areas represent the variations of the GENE turbulent fluxes in the last five GENE-Tango iterations.}
\label{fig:fig_final_es}
\end{figure*}
The electrostatic simulation converged in 28 iterations, while the electromagnetic case required 30 iterations. This is illustrated in Fig.~\ref{fig:fig_final_es} and Fig.~\ref{fig:fig_final_em}, which show the turbulent fluxes from the last five iterations of the GENE-Tango simulations for the electrostatic and electromagnetic cases, respectively. In each case, the fluxes are compared with the corresponding volume-integrated sources and sinks at each radial position within the simulated domain. 
\begin{figure*}
\begin{center}
\includegraphics[scale=0.35]{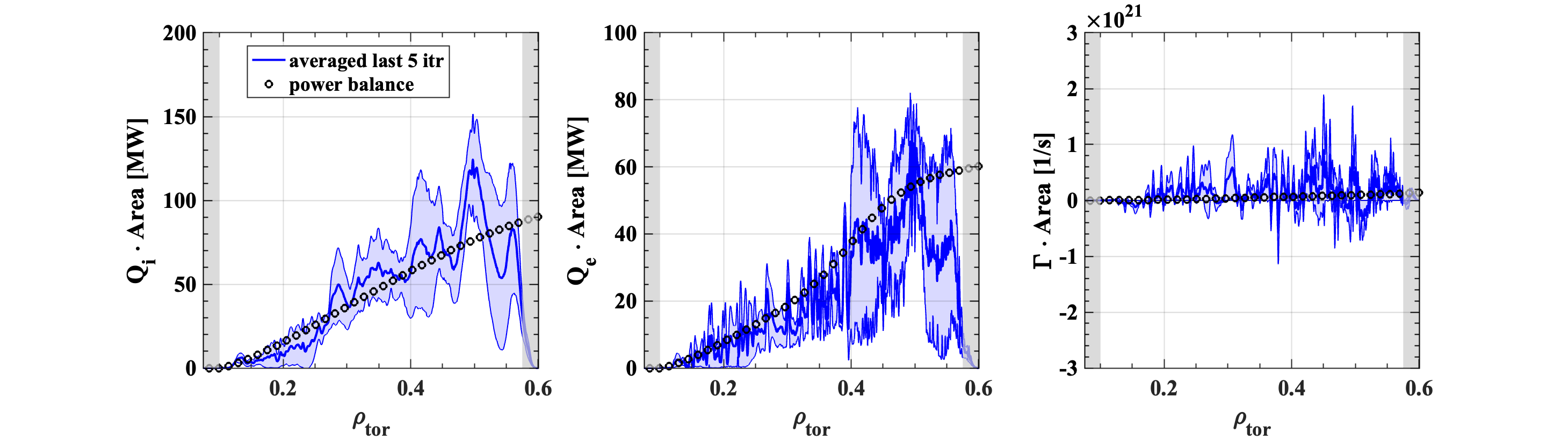}
\par\end{center}
\caption{Time-averaged radial profile of the (a) ion, (b) electron heat fluxes in MW and (c) particle flux in $1/s$ corresponding to the averaged last five GENE–Tango iterations (blue) in the electromagnetic GENE-Tango simulation. The shaded gray areas denote the buffer regions and the black circles the volume integral of the injected particle and heat sources. The shaded blue areas represent the variations of the GENE turbulent fluxes in the last five GENE-Tango iterations.}
\label{fig:fig_final_em}
\end{figure*}
The results show good agreement across all transport channels, indicating that a steady-state solution has been successfully reached. Considering that each iteration spans a simulation time of $t=150 c_s/a$, this corresponds to a total simulated physical time of approximately 12ms and 14ms, respectively. This is about 190 times shorter than a single energy confinement time for the ITER 15 MA baseline scenario, demonstrating that the GENE-Tango framework achieves convergence more than two orders of magnitude faster than traditional flux-driven simulations, which typically require at least one confinement time. This substantial speed-up makes first-principles-based profile prediction for ITER scenarios computationally feasible with global gyrokinetic simulations for the first time.

\begin{figure*}
\begin{center}
\includegraphics[scale=0.35]{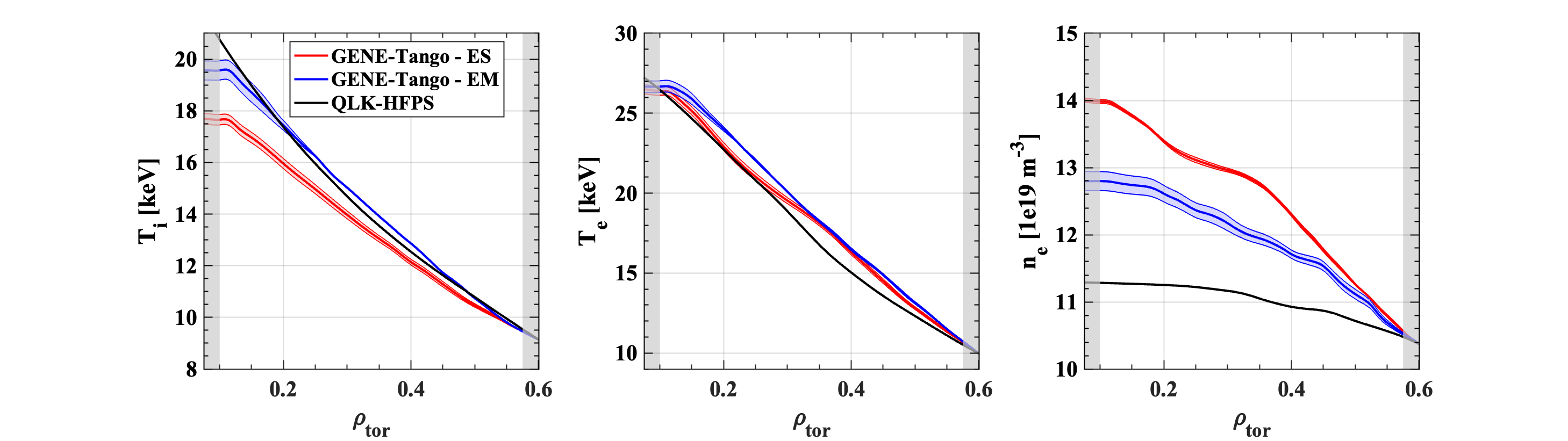}
\par\end{center}
\caption{Comparison of the (a) ion temperature, (b) electron temperature, and (c) density profiles obtained from GENE–Tango simulations in the electrostatic (red) and electromagnetic (blue) cases. The shaded gray areas indicate the buffer regions used in the GENE simulations. The black profiles denote the reference ones obtained by HFPS-QuaLiKiz.}
\label{fig:fig_QLK}
\end{figure*}

Fig.~\ref{fig:fig_QLK} shows the resulting temperature and density profiles for each species obtained with GENE-Tango in both the electrostatic and electromagnetic regimes. 
One key observation is that the inclusion of electromagnetic effects does not significantly impact the electron temperature profile, which remains relatively unchanged across all cases. In contrast, the ion temperature shows larger differences. In particular, it increases from approximately $T_i = 18$keV at the left boundary of the GENE domain in the electrostatic case to $T_i = 20$keV when electromagnetic effects are included. This occurs despite the fact that the electrostatic simulation yields a more peaked plasma density profile, which would typically enhance density-gradient stabilization of ITG turbulence \cite{Romanelli_PF_1989}. The increased ion temperature in the electromagnetic case can be attributed to finite-beta and Shafranov-shift stabilization of ITG turbulence \cite{Bourdelle_NF_2005}, which is particularly effective in the deep plasma core where beta is higher. The electrostatic GENE-Tango simulation produces a much more peaked density profile, reaching $n_e \approx 1.4e20 m^{-3}$. This strong peaking results from a large inward particle pinch due to thermodiffusion driven by a dominant ITG turbulence, thermodiffusion is further reinforced by inward compressibility expected in presence of a relaxed $q$ profile \cite{Bourdelle_PoP_2007, Angioni_PPCF2009}. When electromagnetic effects are included, the density profile predicted by GENE-Tango becomes less peaked. This behavior is consistent with the theoretical findings of Ref.~\cite{Hein_PoP_2010}, which show that electromagnetic fluctuations enhance the outward particle flux contribution of passing electrons in ITG/TEM-dominated regimes. In order to satisfy the particle balance Tango flattens the density profile accordingly.

The density peaking predicted by the GENE-Tango simulations can be compared with the empirical scalings derived from H-mode discharges across multiple devices, as presented in Ref.~\cite{Angioni_PoP_2007}. For the ITER 15 MA baseline scenario analyzed here, the effective collisionality is calculated as $\nu_{\rm eff} = 0.121$ in the electrostatic simulation and $\nu_{\rm eff} = 0.116$ in the electromagnetic one, using the expression $\nu_{\rm eff} = 0.2 \langle n_e \rangle R / \langle T_e \rangle^2$, where $n_e$ is the electron density expressed in units of $1e19m^{-3}$, $R$ is the major radius in meters, $T_e$ is the electron temperature in keV and $\langle \rangle$ represents the volume average. In this context, the density peaking, quantified as the ratio $n_e (\rho_{tor} = 0.2)/\langle n_e \rangle$, is found to be 1.42 in the electrostatic GENE-Tango simulation and 1.31 in the electromagnetic case. This reduction in peaking due to electromagnetic effects aligns with the trends reported in Ref.~\cite{Angioni_PoP_2007}, where increasing electromagnetic influence is associated with weaker inward particle transport. The density peaking predicted in this work has been benchmarked against the scaling laws established by Angioni et al. in Ref.~\cite{Angioni_PoP_2007} from a multi-machine H-mode database comprising Alcator C-Mod, ASDEX Upgrade, and JET. Our results fall within the database at the collisionalities considered toward the bottom of the observed scattered data points. The obtained values are also consistent with CGYRO-Portals predictions for the ITER 15 MA baseline scenario, despite minor differences in the safety factor profile and heating scheme between the scenarios \cite{Howard_NF_2025}. These results suggest that the predicted density profiles fall within the expected range of experimental and theoretical scalings for the given collisionality regime.

\begin{figure}
\begin{center}
\includegraphics[scale=0.25]{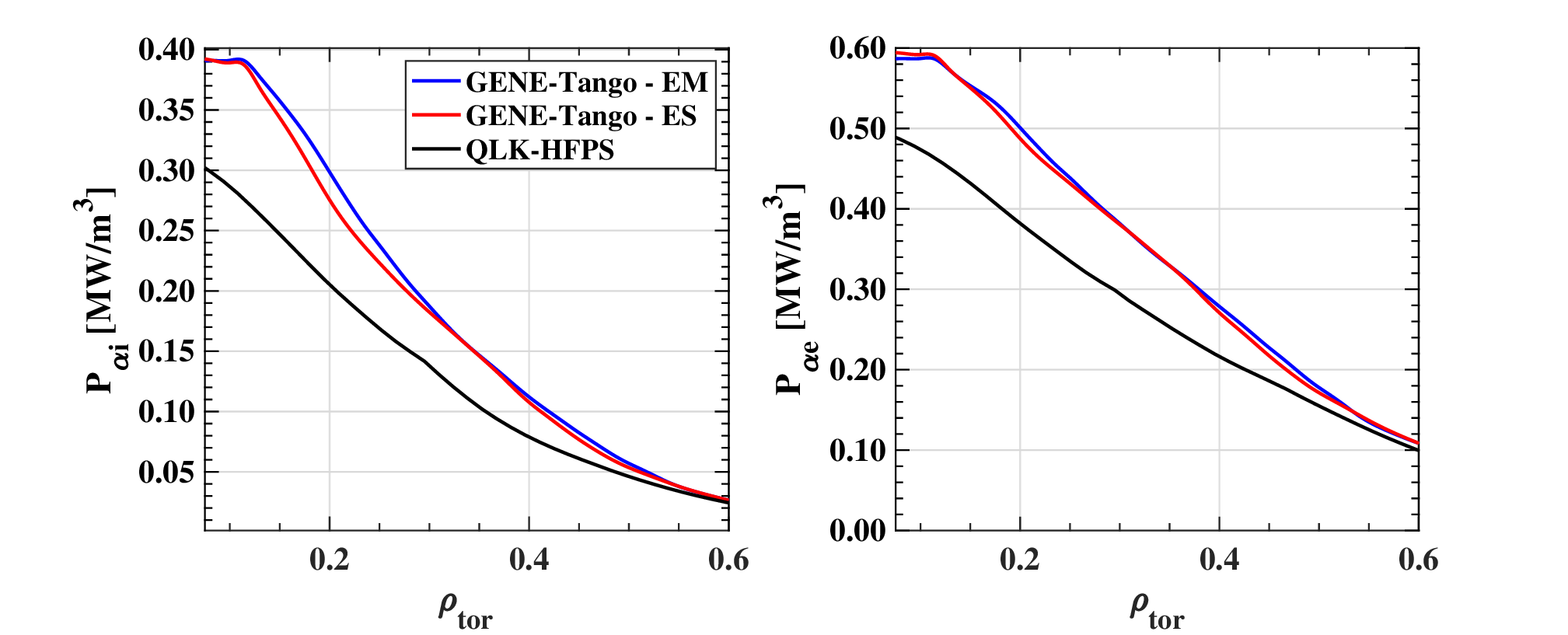}
\par\end{center}
\caption{Comparison of alpha heating power deposited on (a) ions and (b) electrons from GENE–Tango simulations in the electrostatic (red) and electromagnetic (blue) cases, along with results based on initial QuaLiKiz-HFPS profiles (black).}
\label{fig:fig_alpha}
\end{figure}

Interestingly, when evaluating the total alpha heating across the different cases analyzed in this study in Fig.~\ref{fig:fig_alpha}, we find that, despite the differences in the plasma profiles, the GENE-Tango electrostatic and electromagnetic simulations yield very similar fusion power deposited on ions and electrons. 
The integrated alpha heating at the outer boundary of the simulated domain is $P_\alpha \approx 130$MW, corresponding to a total fusion gain of $Q = 12.2$, which exceeds the ITER target of $Q = 10$. This value is obtained by assuming a total injected power of $P_{\text{in}} = 53$MW. In contrast, QualiKiz-HFPS predicts a lower alpha heating power of $P_\alpha = 90$MW, resulting in a fusion gain of $Q =8.5$.

%
%

\section{Characterization of instabilities driving turbulent transport} \label{sec5}

\begin{figure}
\begin{center}
\includegraphics[scale=0.25]{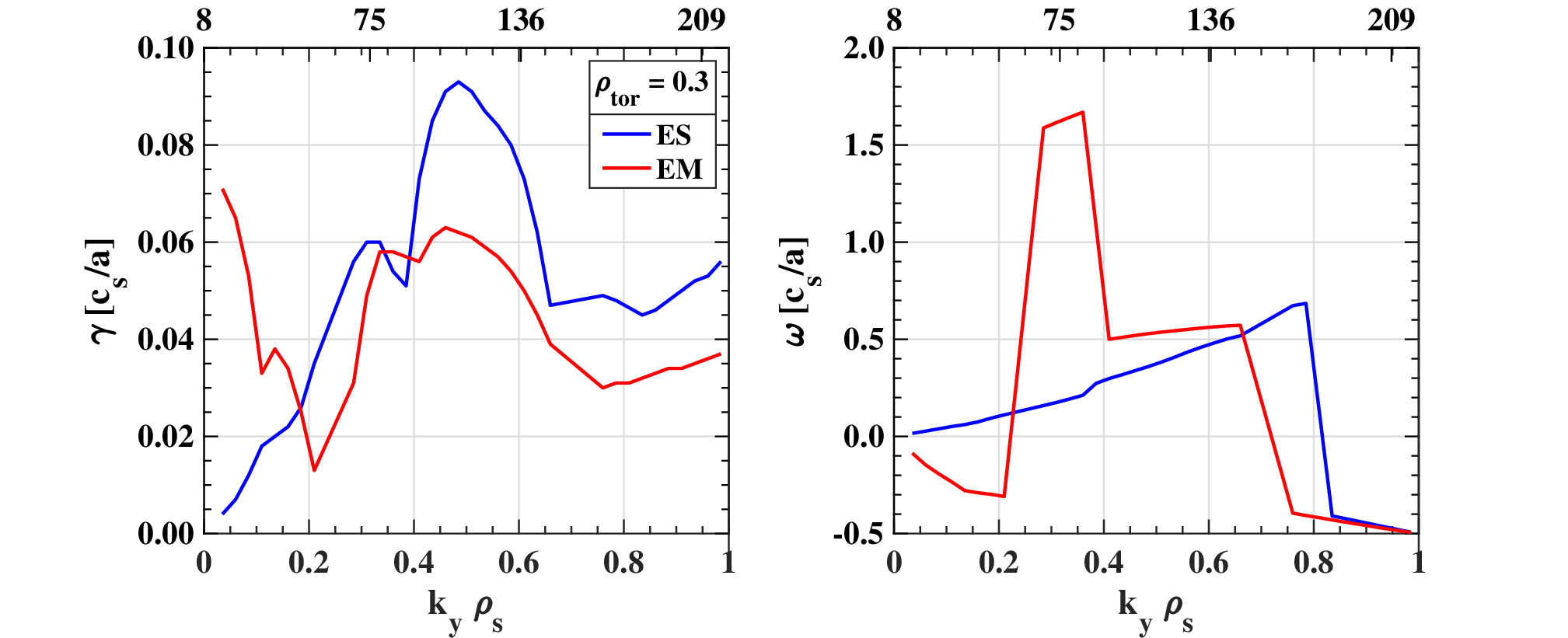}
\par\end{center}
\caption{Comparison of dominant linear growth rates (a) and real frequencies (b) obtained from electrostatic (blue) and electromagnetic (red) flux-tube GENE simulations at $\rho_{tor} = 0.3$ across different toroidal mode numbers. All simulations use the thermal profiles obtained from the final state of the global electromagnetic GENE-Tango simulation.}
\label{fig:fig_linear03}
\end{figure}

\begin{figure}
\begin{center}
\includegraphics[scale=0.25]{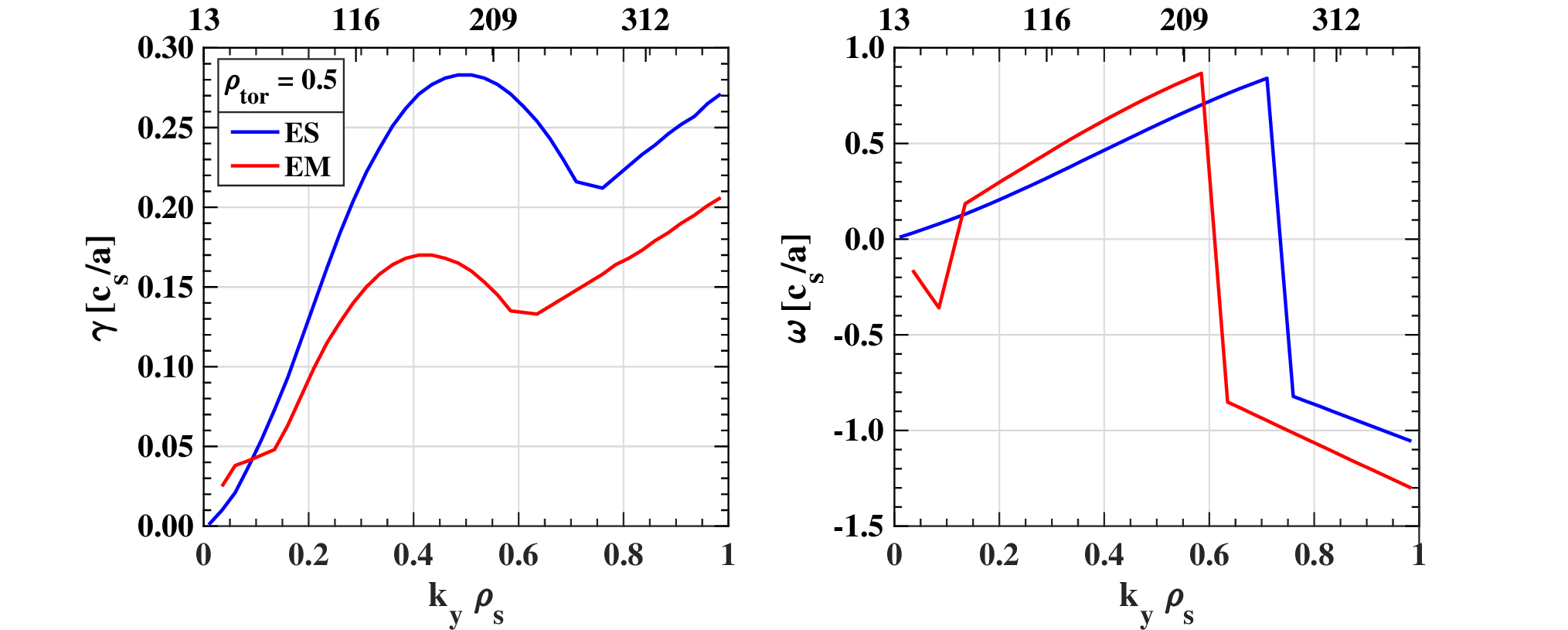}
\par\end{center}
\caption{Comparison of dominant linear growth rates (a) and real frequencies (b) obtained from electrostatic (blue) and electromagnetic (red) flux-tube GENE simulations at $\rho_{tor} = 0.5$ across different toroidal mode numbers. All simulations use the thermal profiles obtained from the final state of the global electromagnetic GENE-Tango simulation.}
\label{fig:fig_linear_05}
\end{figure}

In this section, we identify the dominant instabilities responsible for turbulent transport observed in the nonlinear simulations, for both electrostatic and electromagnetic cases. To this end, we perform linear flux-tube simulations using the final profiles from the electromagnetic GENE-Tango simulation at two radial locations $\rho_{tor} = 0.3$ and $\rho_{tor} = 0.5$. These positions are chosen to be representative, one near the center of the simulation domain and the other closer to the outer region, yet sufficiently far from the global buffer zones. The results for both dominant growth rates and frequencies are presented in Fig.~\ref{fig:fig_linear03} for $\rho_{tor} = 0.3$ and in Fig.~\ref{fig:fig_linear_05} for $\rho_{tor} = 0.5$. 
Focusing first on $\rho_{tor} = 0.3$, we observe a range of instabilities across different $k_y \rho_s$ scales. At low wavenumbers ($k_y \rho_s < 0.2$), an electromagnetic mode is present with frequency in the electron diamagnetic direction.
Interestingly, this EM mode exhibit growth rates that exceed those of ITG modes at higher wave-numbers. In the range $k_y \rho_s = [0.2 - 0.3]$, we find high-frequency modes identified as Alfv\'enic Ion Temperature Gradient (AITG) modes/KBMs \cite{Zonca_1998}, likely destabilized by the strong ion temperature gradient in the core. At higher $k_y \rho_s$, ITGs dominate, followed by TEMs at even smaller scales. 

When performing electrostatic simulations using the same background profiles as those used in the linear electromagnetic simulations discussed above, we observe minimal changes in the ITG and TEM instabilities, which retain similar frequencies and exhibit slightly increased growth rates. As expected, however, both the EM modes and AITGs/KBMs at low wavenumbers are completely stabilized in the absence of electromagnetic effects.

At the radial location $\rho_{tor} = 0.5$, we observe ITGs and TEMs as the dominant instabilities in the electromagnetic simulations, along with a weak MTM branch destabilized at low values of $k_y \rho_s$, characterized by odd parity in the electrostatic potential and even parity in the parallel magnetic vector potential (not shown) \cite{Hazeltine_PF_1975,Doerk_PRL_2011,Hatch_NF_2015}. However, these MTMs exhibit very low growth rates, only slightly exceeding those of the ITG modes. Similarly to the case at $\rho_{tor} = 0.3$, when performing electrostatic simulations using the same background profiles as in the electromagnetic case, the MTM branch disappears, and the growth rates of ITGs and TEMs increase due to the absence of electromagnetic stabilization.

\begin{figure*}
\begin{center}
\includegraphics[scale=0.45]{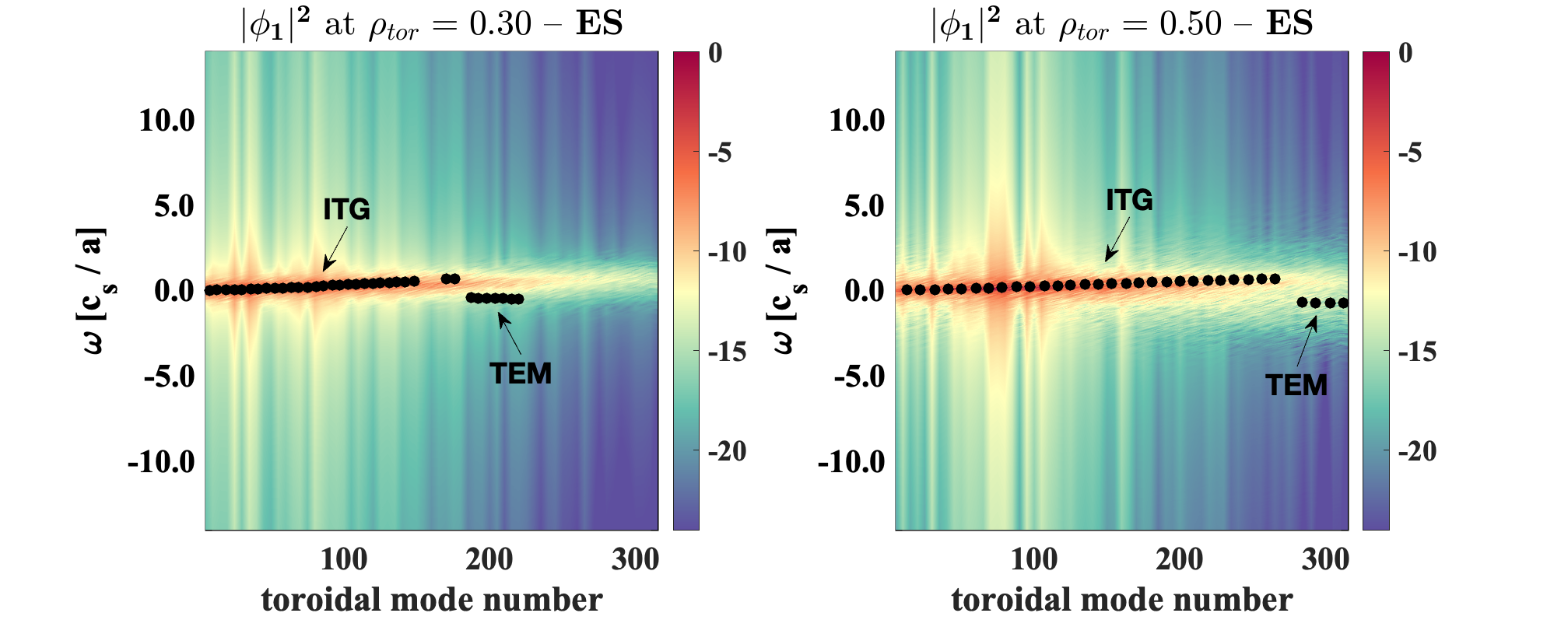}
\includegraphics[scale=0.45]{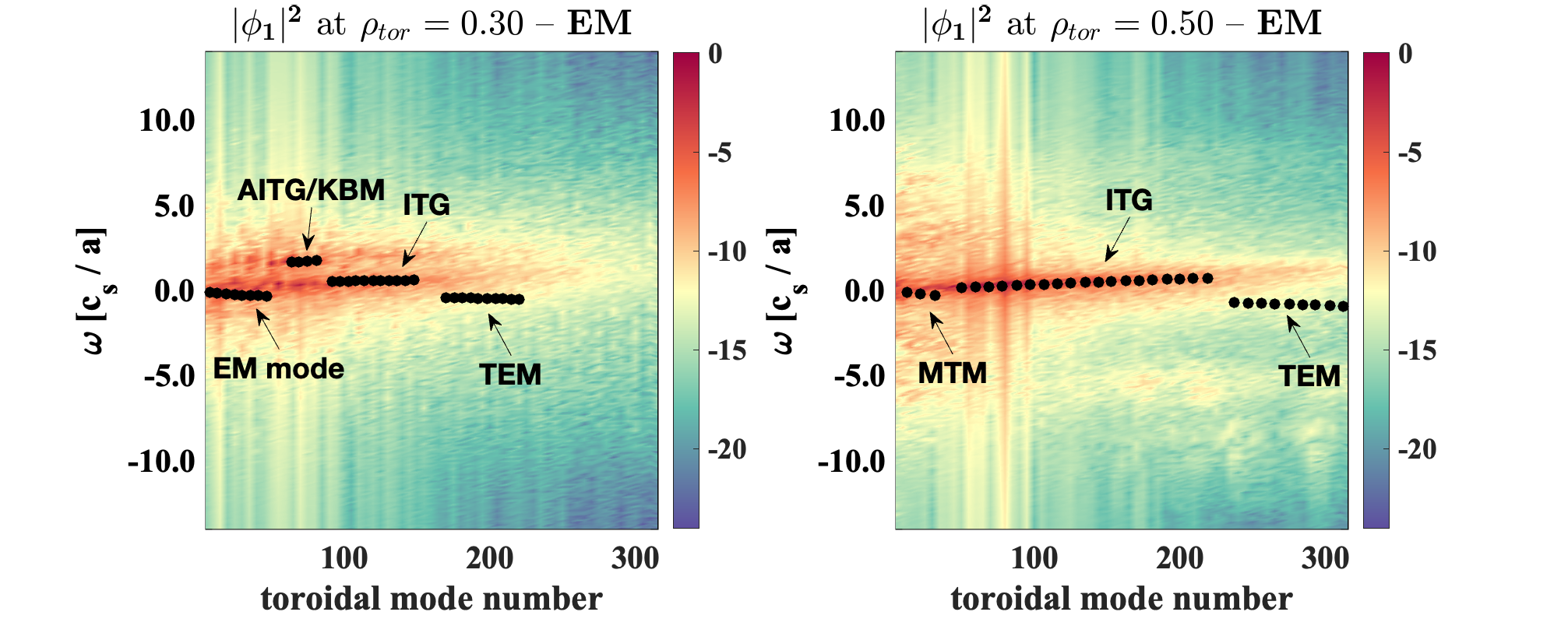}
\par\end{center}
\caption{Frequency spectra of the electrostatic potential $|\phi_1|^2$ as a function of the toroidal mode number at radial locations $\rho_{tor} = 0.3$ ((a), (c)) and $\rho_{tor} = 0.5$ ((b), (d)), from the electrostatic (top panels) and electromagnetic (bottom panels) global GENE simulations using the corresponding steady-state profiles. The spectra are computed over the time window $t [c_s/a] = [500 - 750]$, averaged along the field-aligned coordinate z, and corrected for the Doppler shift due to toroidal plasma rotation. The amplitude is shown on a logarithmic scale. Black dots indicate the linear mode frequencies obtained from the flux-tube simulations (Fig.~\ref{fig:fig_linear03} and Fig.~\ref{fig:fig_linear_05}) at each radial location and toroidal mode number.}
\label{fig:fig_phi}
\end{figure*}

\begin{figure}
\begin{center}
\includegraphics[scale=0.25]{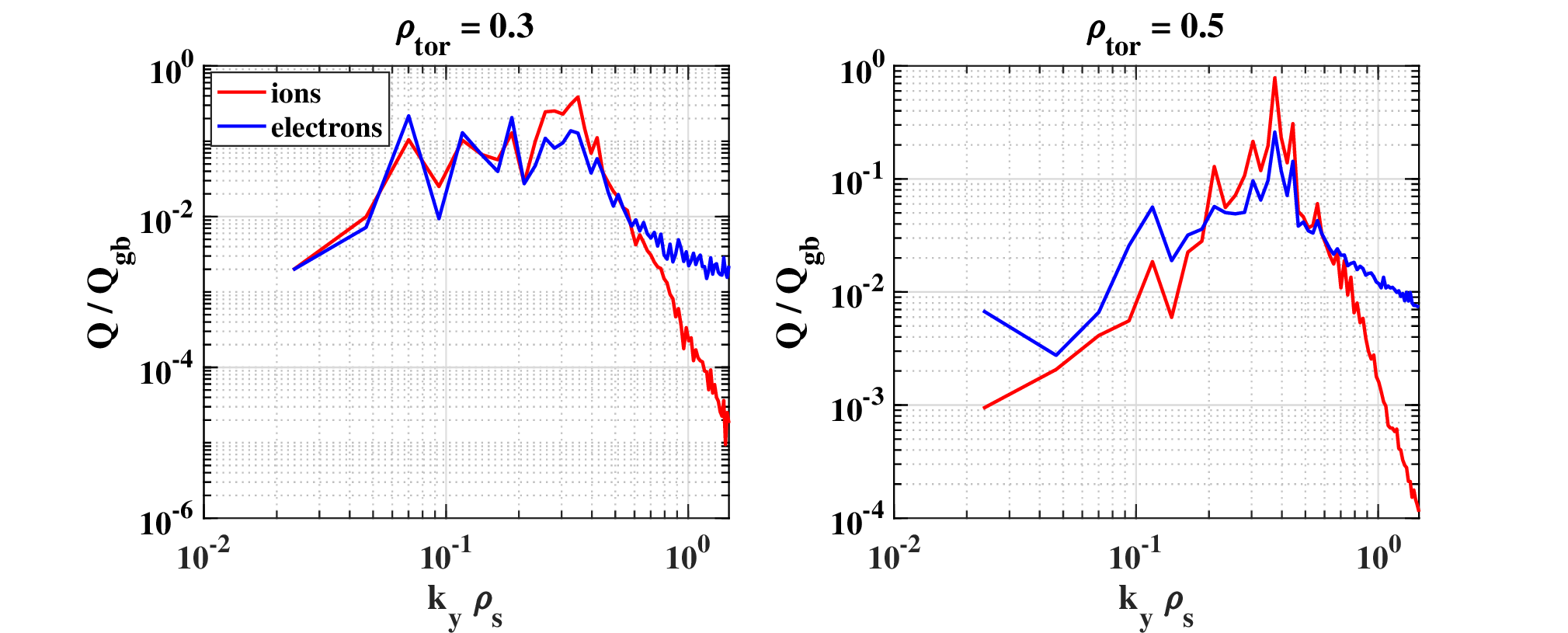}
\par\end{center}
\caption{Nonlinear time-averaged heat flux spectra from the global electromagnetic GENE simulation using the steady-state electromagnetic profiles obtained from the GENE-Tango run, at the radial locations $\rho_{tor} = 0.3$ (a) and $\rho_{tor} = 0.5$ (b). Ion and electron heat fluxes are shown in red and blue, respectively. The spectra are averaged over the nonlinear saturated phase.}
\label{fig:fig_spectra}
\end{figure}

The plasma instabilities identified through linear simulations at the two considered radial locations are also observed in the frequency spectra of the electrostatic potential extracted from the global nonlinear GENE simulations, both in the electrostatic and electromagnetic cases. This is shown in Fig.~\ref{fig:fig_phi}.
In particular, when using the final GENE-Tango profiles in the electrostatic limit (Fig.~\ref{fig:fig_phi} top panels), the signal is entirely dominated by ITG modes, with an excellent agreement between the linear and nonlinear frequencies. A slight mismatch is observed for the TEM branch. However, this occurs at higher toroidal mode numbers (or, equivalently, at larger $k_y \rho_s$), which do not contribute significantly to the turbulent transport (as shown in Fig.~\ref{fig:fig_spectra}). No major qualitative differences are observed between the nonlinear frequency spectra at $\rho_{tor} = 0.3$ and $\rho_{tor} = 0.5$ in the electrostatic case. 

In contrast, the nonlinear frequency spectra from the global electromagnetic GENE simulation at $\rho_{tor} = 0.3$, using the final GENE-Tango electromagnetic profiles (Fig.~\ref{fig:fig_phi} bottom panels), shows a significantly broader frequency band compared to the electrostatic case. While ITG modes still contribute, a pronounced branch appears at lower toroidal mode numbers, corresponding to the frequencies of the AITG/KBMs modes identified in the linear flux-tube simulations at $\rho_{tor} = 0.3$. This indicates that AITG/KBM modes are indeed unstable in the global nonlinear simulation, with similar frequency as observed in the linear simulations. Additionally, a broad band is observed at even lower toroidal mode numbers, consistent with the linear signature of the electromagnetic mode found in the linear simulations at low wavenumbers, suggesting that this mode may also persist in the global nonlinear electromagnetic simulations. At the radial location $\rho_{tor} = 0.5$, the frequency spectrum remains consistent with the linear results, which closely reproduce the features observed in the nonlinear simulation. In particular, a weak branch appears at scales corresponding to those of the MTM modes identified in the linear analysis, suggesting their possible presence also in the nonlinear regime. At this radial location, the TEM branch is again not observed in the global nonlinear spectra. However, this is expected, as these modes correspond to scales that do not contribute significantly to turbulent transport. 
This is illustrated in Fig.~\ref{fig:fig_spectra}, where the ion and electron heat flux spectra from the electromagnetic simulation are shown at the two selected radial positions. The figure shows that, at the scales where TEMs are linearly unstable, the turbulent flux is more than one order of magnitude smaller than that at ITG scales for electrons, and over five orders of magnitude smaller for ions. Moreover, Fig.~\ref{fig:fig_spectra} reveals distinct peaks in both electron and ion heat flux at $\rho_{tor} = 0.3$ in the range $k_y \rho_s < 0.2$, which corresponds to the scales where the electromagnetic modes were identified in the linear simulations. Similarly, additional peaks are observed in the range $k_y \rho_s = [0.2 - 0.3]$, matching the scales of the AITG/KBM modes. These findings indicate that both the EM mode at low wavenumbers and AITGs/KBMs contribute non-negligibly to the thermal transport in the global GENE simulations and should not be neglected in transport modelling. Finally, the spectra at $\rho_{tor} = 0.5$ also exhibit features consistent with the linear results, further supporting the presence of weak but detectable MTM signatures at this location.

\section{Comparison with flux-tube GENE simulations} \label{sec6}

Flux-tube (local) gyrokinetic simulations are widely used to model turbulent transport across a broad range of plasma conditions and are generally expected to yield results comparable to those of more computationally demanding global simulations in large-scale devices like ITER, where global effects are not expected to play a significant role \cite{McMillan_PRL_2010}. In this section, we compare the global turbulent fluxes obtained from the electromagnetic GENE simulation using the final steady-state profiles with those from flux-tube simulations performed at two radial positions, namely $\rho_{tor} = 0.3$ and $\rho_{tor} = 0.5$.

These simulations are carried out using a spatial resolution of $(n_x \times n_{k_y} \times n_z) = (384 \times 96 \times 32)$, corresponding to the radial ($x$), bi-normal ($y$), and field-aligned ($z$) directions, respectively. The box size in the radial and bi-normal directions is approximately $L_x \approx 156$ and $L_y \approx 782$ in GENE normalized units. The velocity space is resolved using $(n_{v_\shortparallel} \times n_\mu) = (32 \times 16)$ grid points, with domain extents of 3 and 9 in the parallel velocity and magnetic moment directions, respectively. The minimum bi-normal wave number is set to $k_{y,\text{min}} \rho_s \approx 0.013$ (corresponding to the toroidal mode number $n = 6$), and the maximum value reaches $k_{y,\text{max}} \rho_s \approx 1.23$. These grid resolutions were determined based on convergence studies.

\begin{figure*}
\begin{center}
\includegraphics[scale=0.35]{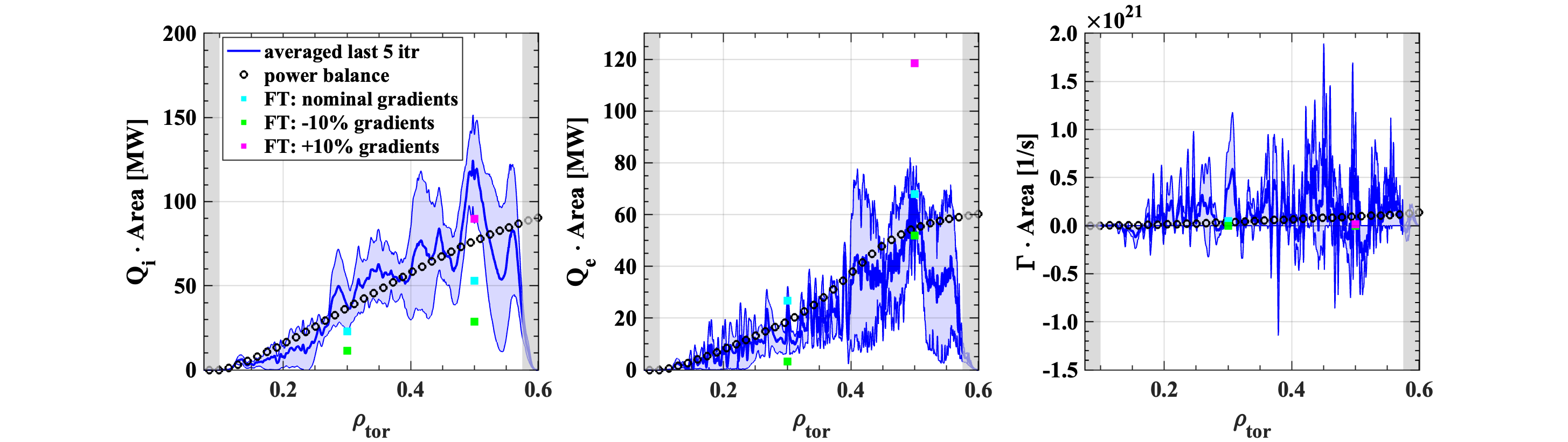}
\par\end{center}
\caption{Time-averaged radial profiles of (a) ion heat flux, (b) electron heat flux (both in MW), and (c) particle flux (in $1/s$) corresponding to the average over the last five GENE–Tango iterations (blue) from the electromagnetic GENE-Tango simulation. Also shown are flux-tube simulation results at the radial locations $\rho_{tor} = 0.3$ and $\rho_{tor} = 0.5$, using the nominal gradients from the steady-state electromagnetic GENE-Tango profiles (cyan dots) and with gradient variations of $\pm 10\%$ (magenta and green dots, respectively). The flux-tube simulation at $\rho_{tor} = 0.3$ with a $10\%$ increase in gradients did not converge, producing fluxes above 500 MW for both ions and electrons, and is therefore excluded. Gray shaded areas indicate buffer regions, while blue shaded areas represent the range of turbulent flux variations over the last five GENE-Tango iterations.}
\label{fig:fig_localglobal}
\end{figure*}

The resulting turbulent fluxes for each channel are summarized in Fig.~\ref{fig:fig_localglobal} (cyan squares), corresponding to the nominal gradients taken from the final GENE-Tango profiles. 
Fig.~\ref{fig:fig_localglobal} shows that the particle flux is generally well reproduced and remains close to the target flux determined by the volume-integrated external sources for both radial locations. Regarding the heat fluxes, we observe a consistent tendency to overestimate the electron heat flux and underestimate the ion heat flux. Nevertheless, both remain reasonably close to the target heat flux and to the global simulation results.

To assess the stiffness of turbulent transport in the ITER baseline scenario, we perform additional flux-tube simulations at the two selected radial locations by changing the ion and electron temperature and density gradients by $10\%$, while keeping all other parameters fixed. The results, summarized in Fig.~\ref{fig:fig_localglobal}, reveal that transport is extremely stiff and even a change of $10\%$ in the gradients can lead to variations in the heat fluxes of up to $100\%$ for both ions and electrons. Interestingly, we observe (not shown here) that the electromagnetic component of the electron heat flux strongly increases as the gradients are increased. This is linked to the progressive destabilization of unstable MTMs, EM modes at low wavenumber and KBMs/AITGs.

These findings support the interpretation that the final electromagnetic profiles obtained from the stand-alone GENE-Tango simulation lie in a parameter regime where MTMs and KBMs/AITGs are weakly unstable. In this regime, MTMs can survive in nonlinear simulations and drive finite electron heat transport with negligible particle transport, thereby allowing a good match with the target fluxes. However, based on fingerprint analysis \cite{Kotschenreuther_NF_2019}, MTMs alone are insufficient to account for the total ion-scale transport, indicating the presence of a coexisting ITG turbulence component to provide the necessary ion heat flux. However, if the gradient drive becomes too large, the MTM-driven electron transport increases sharply, more rapidly than the ion heat or particle fluxes, causing the total fluxes to deviate significantly from the target values. A similar effect is observed with the KBMs/AITGs, which, when destabilized, lead to an increase in the electron heat flux, particularly its electromagnetic component \cite{DiSiena_NF_2023_2}.
\begin{figure}
\begin{center}
\includegraphics[scale=0.25]{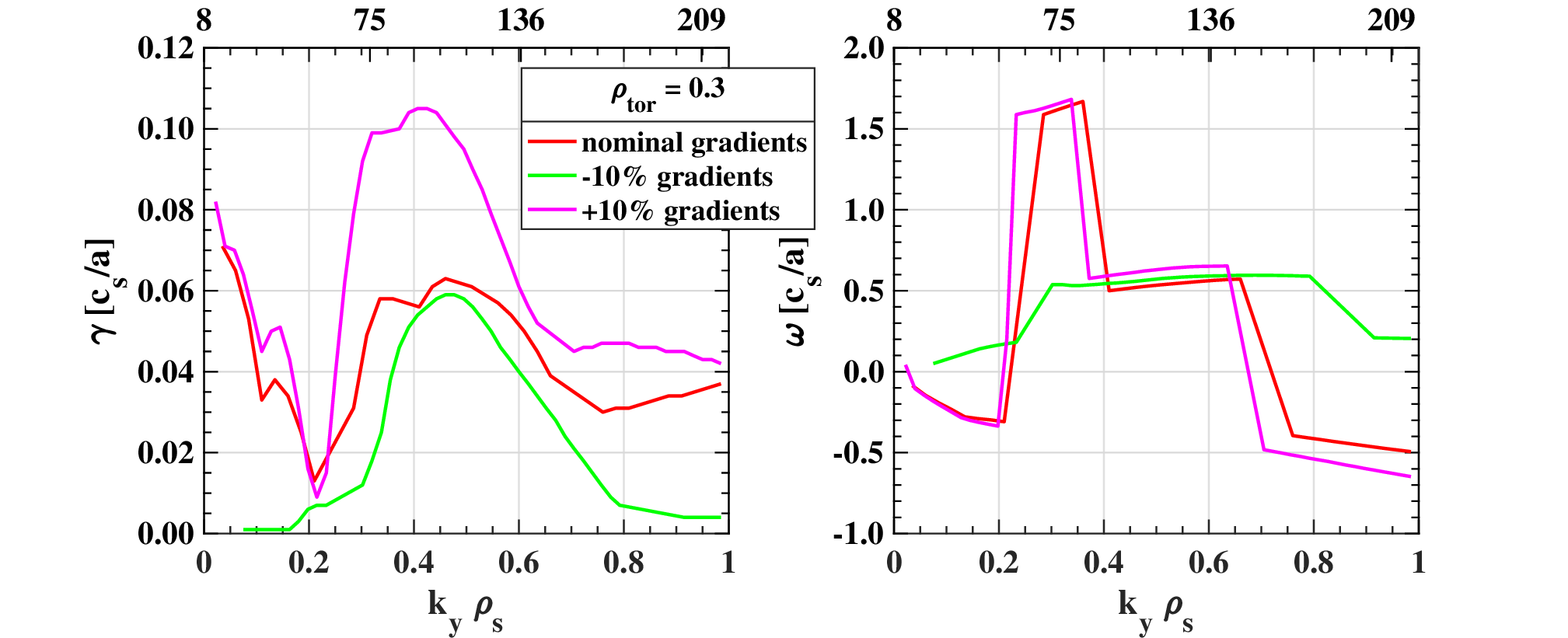}
\par\end{center}
\caption{Comparison of dominant linear growth rates (a) and real frequencies (b) from electromagnetic flux-tube GENE simulations at $\rho_{tor} = 0.3$, performed across a range of toroidal mode numbers. Results are shown for the nominal gradient profiles obtained from the steady-state electromagnetic GENE-Tango simulation (red), as well as for cases with ion and electron temperature and density gradients increased (magenta) and decreased by $10\%$ (green), respectively.}
\label{fig:fig_linear_10}
\end{figure}
This interpretation is further supported by the linear analysis shown in Fig.~\ref{fig:fig_linear_10}, which illustrates the growth rates and mode frequencies at $\rho_{\text{tor}} = 0.3$ for both the nominal and $\pm 10\%$ gradient cases. The results shows that even a $10\%$ change in the gradients leads to a substantial increase in the AITG/KBM growth rates, while the growth rates of the EM mode at low wavenumber exhibit a more moderate increase. The presence of AITGs/KBMs and MTMs (at $\rho_{tor} = 0.5$) in these local GENE simulations suggests that a global gyrokinetic approach is likely more accurate than a flux-tube one in this regime. This is because both AITGs/KBMs and MTMs are particularly sensitive to the safety factor profile, and its numerical linearization in flux-tube simulations may significantly affect the mode drive—e.g., by shifting the position of rational surfaces or introducing artificial rational surfaces not present in the scenario. These effects are properly captured only in radially global simulations. Nevertheless, since the steady-state profiles only weakly excite these modes, flux-tube simulations can still provide qualitatively consistent predictions with the global results, as shown in Fig.~\ref{fig:fig_localglobal}.

\section{Comparison of GENE-Tango predicted profiles with HFPS-QuaLiKiz and HFPS-TGLF} \label{sec7}

Using the same profiles as fixed parameters, with boundary conditions at $\rho_{tor} = 0.6$ and initial conditions from GENE-Tango, the electron and ion heat fluxes, as well as the D and T particle fluxes, are computed within the High Fidelity Plasma Simulator (HFPS, an IMAS-compatible version of JINTRAC). This integrated modeling framework then predicts the $T_i$, $T_e$ and $n_e$ profiles up to $\rho_{tor} = 0.6$. QuaLiKiz, which is electrostatic, and TGLF with saturation rule 2, capable of modeling both electrostatic and electromagnetic turbulence, are employed in this study. In particular, the settings used for TGLFsat2 electromagnetic are based on recent validations against the linear gyrokinetic code GKW for JET high-$\beta$ plasmas \cite{Najlaoui_2025}, which have been further improved. Namely, for TGLFsat2 electromagnetic the following settings are used: $\text{kygrid}_{\text{model}} = 4$, $\text{alpha}_{\text{zf}}=-1$, $\text{width}=3$, $\text{filter}=-0.1$, $\text{min}_{\text{width}}=-0.3$, $\text{nbasis}= 6$, $\text{use}_{{\text{MHD}}_{\text{rule}}}=true$, $\text{use}_{\text{bperp}}=true$ and $\text{use}_{\text{bpar}}=false$ (see \cite{Najlaoui_2025} for the meaning of these code parameters). TGLFsat2 electrostatic is used with the default settings and without magnetic fluctuations, namely $\text{kygrid}_{\text{model}}=4$, $\text{alpha}_{\text{zf}}=1$, $\text{width}=1.65$, $\text{filter}=2$, $\text{min}_{\text{width}}=0.3$, $\text{nbasis}= 6$, $\text{use}_{{\text{MHD}}_{\text{rule}}}=false$, $\text{use}_{\text{bperp}}=false$ and $\text{use}_{\text{bpar}}=false$. The profiles are time averaged over 5 s, after 55 s of plasma evolution in which the $q$, $V_{tor}$ and $Z_{\rm eff}$ profiles are kept fixed as done in GENE-Tango. 
\begin{figure*}
\begin{center}
\includegraphics[scale=0.35]{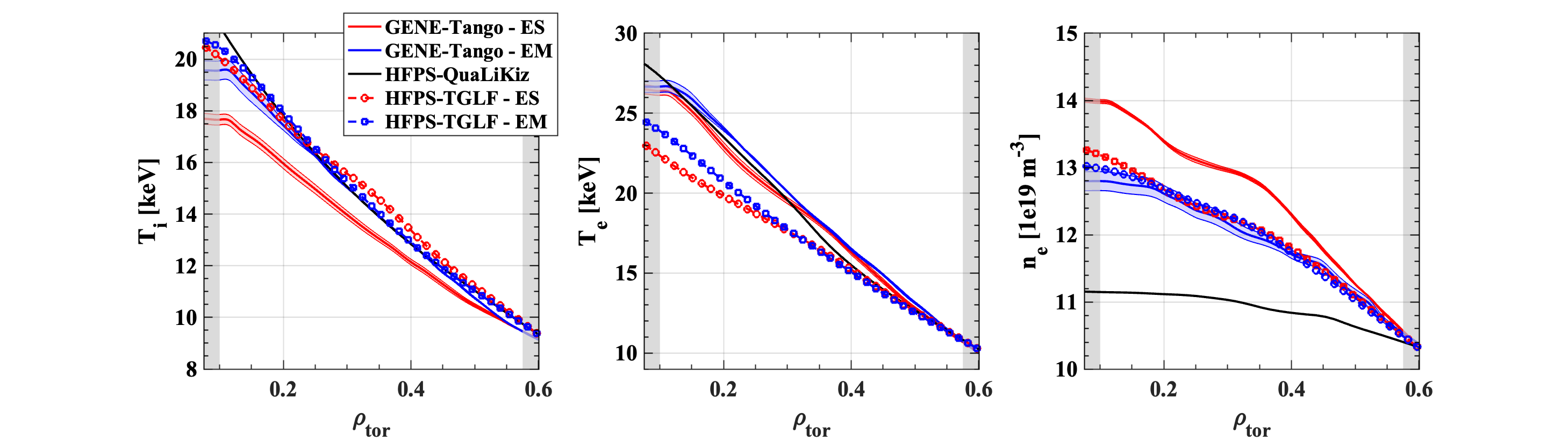}
\par\end{center}
\caption{Comparison of the (a) ion temperature, (b) electron temperature, and (c) density profiles obtained from GENE–Tango simulations in the electrostatic (red) and electromagnetic (blue) cases and obtained from HFPS-QuaLiKiz (black), HFPS-TGLFsat2 electrostatic (dashed, red) and HFPS-TGLFsat2 electromagnetic (dashed, blue).}
\label{fig:Tango_JINTRAC}
\end{figure*}

For the electrostatic predictions, in red and black on Fig.~\ref{fig:Tango_JINTRAC}, the TGLFsat2 predicted profiles are similar to GENE-Tango electrostatic predictions, with $T_i$ slightly over-predicted, $T_e$ slightly under-predicted and $n_e$ slightly under-predicted. The QuaLiKiz predicted temperature profiles also agree rather well with GENE-Tango electrostatic predictions, while the density predicted by QuaLiKiz is much flatter than GENE-Tango. It is also flatter than the QuaLiKiz prediction reported in \cite{Citrin_PPCF_2023}, where the $q$ profile was flatter. A likely cause of discrepancy in the low collisionality ITER core, is due to different ways of speeding up QuaLiKiz and TGLF linear responses. Indeed, these reduced models separate the passing particle from the trapped and simplify their phase-space integration. Such approximation can impact the balance betweeen ITG and TEM and hence the predictions of density profiles. This is documented for QuaLiKiz in \cite{Stephens_2021, Stephens_2024} and for TGLFsat2 in \cite{Staebler_NF_2021}. In TGLFsat2, improved models for the loss of bounce averaging and electron collisions have been implemented. Concerning the electromagnetic prediction by TGLFsat2 (dashed blue lines on Fig.~\ref{fig:Tango_JINTRAC}), the agreement on the $T_e$, $T_i$ and $n_e$ profile predictions when compared to GENE-Tango electromagnetic is very good on $T_i$ and $n_e$ and slightly under-predicted for $T_e$. Stand alone verification of the linear spectra of GENE, GKW, TGLFsat2 and QuaLiKiz will be carried out. The goal will be to verify further TGLFsat2 in high $\beta$ regime from JET case published in \cite{Najlaoui_2025} to JT60-SA and ITER. 

\section{Role of plasma rotation on plasma core turbulence} \label{sec8}

\begin{figure}
\begin{center}
\includegraphics[scale=0.25]{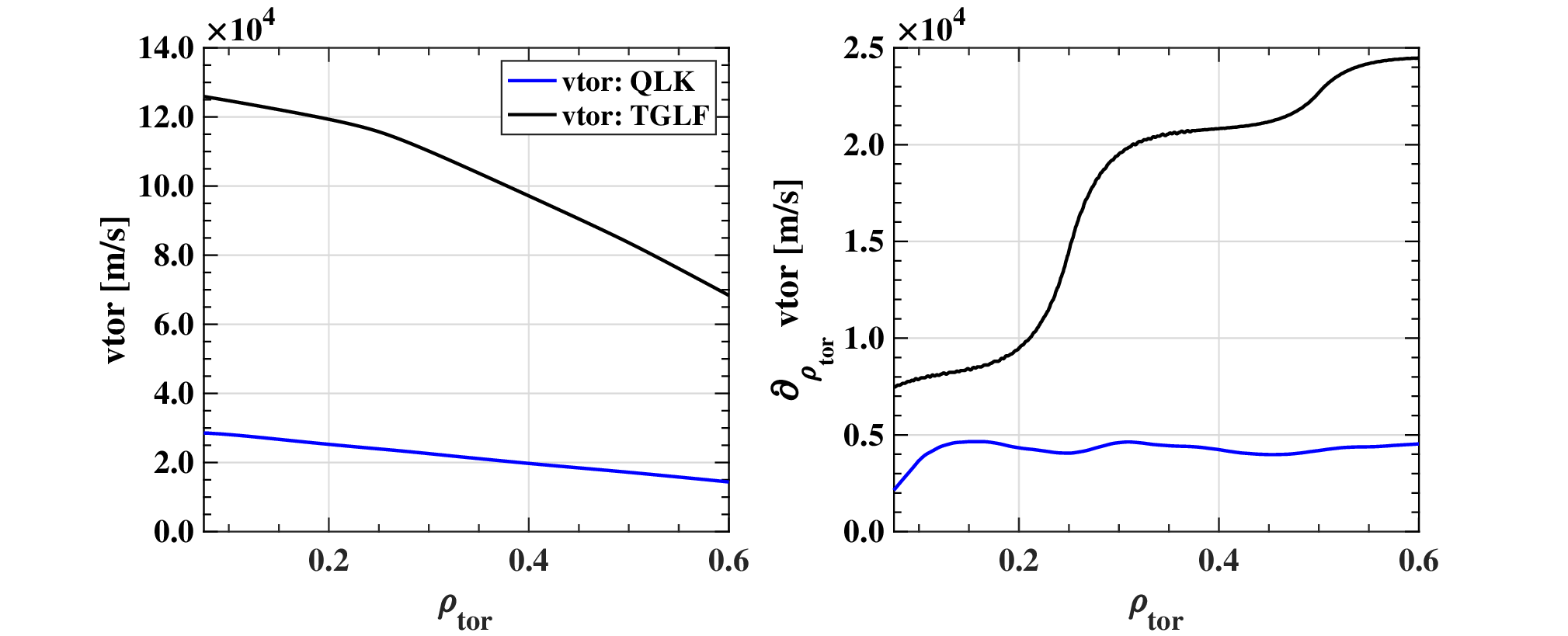}
\par\end{center}
\caption{Comparison of (a) the toroidal rotation profiles predicted based on a Prantdl number of 1 using QuaLiKiz (blue) and based on Chrystal's model \cite{Chrystal_NF_2020} (black), and (b) their corresponding radial derivatives.}
\label{fig:fig_rotation}
\end{figure}

The gyrokinetic simulations presented in this manuscript use the toroidal rotation profile computed using QuaLiKiz with a Prandtl number of 1, see section \ref{sec2}. The toroidal rotation is held fixed, see Fig.~\ref{fig:fig_zeff}, throughout the iterations. Determining the toroidal rotation for future tokamaks like ITER is a challenging problem. Moreover, the toroidal rotation gradient impacts the $E\times B$ shear which itself can impact the turbulence drive, especially at low wave numbers \cite{Burrell_2020}. Therefore, in this section, we perform global, gradient-driven GENE simulations to compare results obtained with the final GENE-Tango profiles, using either the nominal toroidal rotation or an alternative profile using empirical pedestal top scaling and momentum transport in the core \cite{Chrystal_NF_2020} combined in ASTRA and used for ITER predictions \cite{ Luda_NF_2025}. A comparison of the different toroidal rotation profiles is shown in Fig.~\ref{fig:fig_rotation}, along with their radial derivatives, which effectively act as a shearing term in the gyrokinetic equations.

The rotation profile computed based on \cite{Chrystal_NF_2020, Luda_NF_2025} is significantly higher, reaching values of $v_{tor} \approx 120 km/s$, than the one used as a reference \cite{Mantica_PPCF_2020}. Moreover, its radial derivative is also substantially larger than that of the reference profile, implying a stronger stabilizing shear effect.

\begin{figure*}
\begin{center}
\includegraphics[scale=0.35]{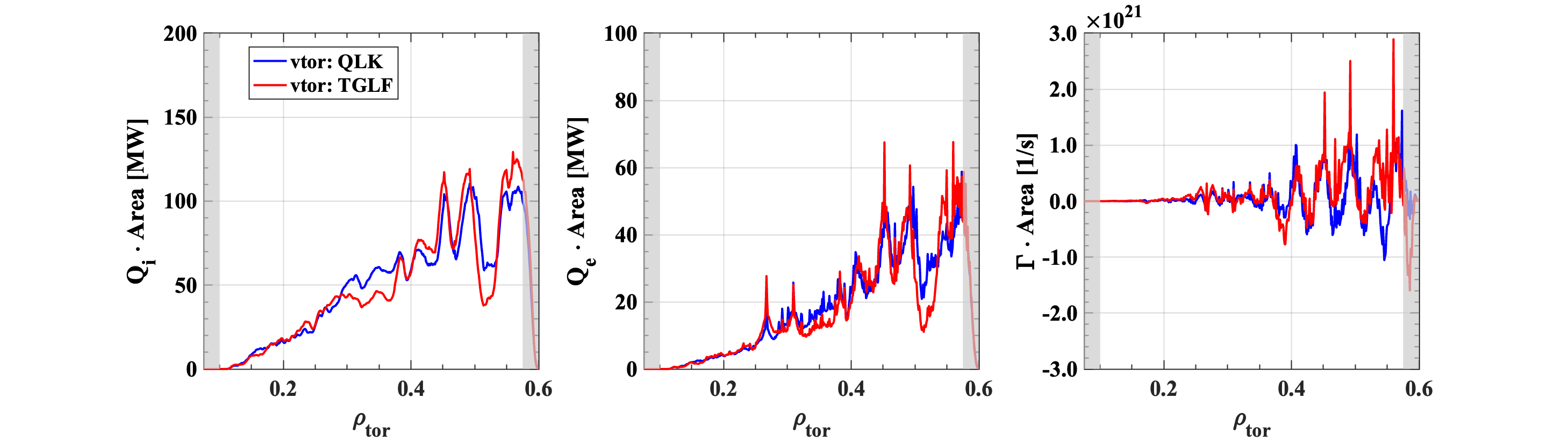}
\par\end{center}
\caption{Comparison of the time-averaged radial profiles of the (a) ion heat flux, (b) electron heat flux (both in MW), and (c) particle flux (in $1/s$) obtained from radially global electrostatic GENE simulations. The simulations use the steady-state profiles from the electrostatic GENE-Tango simulation, employing toroidal rotation profiles based on a Prantdl number of 1 using QuaLiKiz (blue) and based on Chrystal's model \cite{Chrystal_NF_2020} (red). The shaded gray areas denote the buffer regions.}
\label{fig:fig_vtor_flux}
\end{figure*}

Our analysis aims to shed light on whether toroidal rotation can contribute to reducing turbulent transport in the core of the ITER 15MA baseline scenario. These analyses are performed in the electrostatic limit. All simulations use the same numerical resolution as described in Sec.~\ref{sec2}. The results, summarized in Fig.~\ref{fig:fig_vtor_flux}, show time-averaged radial profiles of the turbulent fluxes for both ions (mixed deuterium-tritium species) and electrons, as well as the particle flux.
Fig.~\ref{fig:fig_vtor_flux} shows that increasing both the toroidal rotation and its radial gradient by nearly a factor of four at $\rho_{tor} = 0.6$ does not significantly affect the turbulent fluxes, which remain close to the nominal values obtained using the original toroidal rotation profile. This result can be explained by comparing the zonal flow shearing rates $\omega_{E \times B}$ with the linear growth rates of the most unstable drift-wave modes for the two rotation profiles at $\rho_{tor} = 0.5$. Specifically, $\omega_{E \times B} \approx 0.014$ for the reference and $\omega_{E \times B} \approx 0.058$ for the case based on Chrystal's model \cite{Chrystal_NF_2020}, both of which are small compared to the corresponding linear growth rates by a factor of $12$ and $3$ respectively (see Fig.~\ref{fig:fig_linear_05}). As a result, neither toroidal rotation profile induces a sufficient shearing rate to significantly suppress electrostatic turbulence in the core of this ITER baseline scenario.

\section{Impact of the safety factor profile} \label{sec9}

In the previous sections, we have shown that electromagnetic turbulence plays a significant role in this ITER baseline scenario. On JET high $\beta$ core plasmas, the AITG/KBM are shown to be unstable and favored by a flat q profile \cite{Kumar_2021}. In the case we use here for reference, a relaxed safety factor profile reaching $q < 1$ on axis is used (see \ref{sec2}). Given the strong sensitivity of electromagnetic turbulence to the safety factor profile, we now investigate the impact of a nearly flat $q$ profile slightly above unity, characteristic of a post sawtooth crash. Such a profile is characterized by near-zero magnetic shear in the plasma core. 
\begin{figure}
\begin{center}
\includegraphics[scale=0.30]{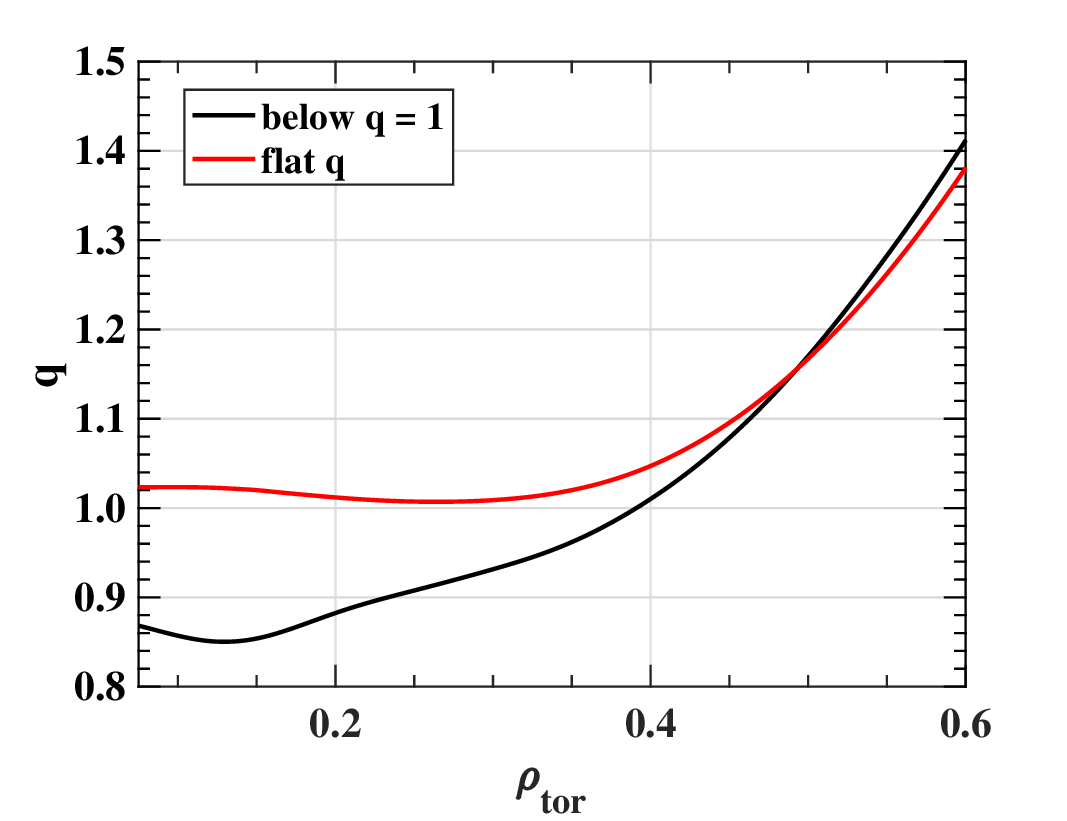}
\par\end{center}
\caption{Comparison of the safety factor radial profile for the ITER 15MA baseline scenario showing the original case where q drops below 1 in the core (black) and a modified case in which the profile is flattened to roughly $q = 1$ in the core (red).}
\label{fig:fig_flatq}
\end{figure}
Fig.~\ref{fig:fig_flatq} compares the safety factor and magnetic shear profiles used in the previous analyses with those considered in this section. The new magnetic geometry, characterized by a flat safety-factor profile slightly above unity in the core, has been reconstructed with CHEASE using the pressure profile from the final iteration of the GENE-Tango electromagnetic simulation. Here, the core pressure profile is not flattened by a sawtooth crash and/or assumed to have recovered, while the current density has not yet fully relaxed owing to the slower timescale of current diffusion. In particular, we perform a gradient-driven electromagnetic global GENE simulation using the final GENE-Tango profile obtained from the electromagnetic case but using the modified safety factor profile. The numerical resolution and simulation setup are identical to those described in Sec.~\ref{sec3}. The results of this comparison are presented in Fig.~\ref{fig:fig_flat_heat}, which shows the time-averaged turbulent fluxes for each species corresponding to the two different safety factor profiles.
\begin{figure*}
\begin{center}
\includegraphics[scale=0.35]{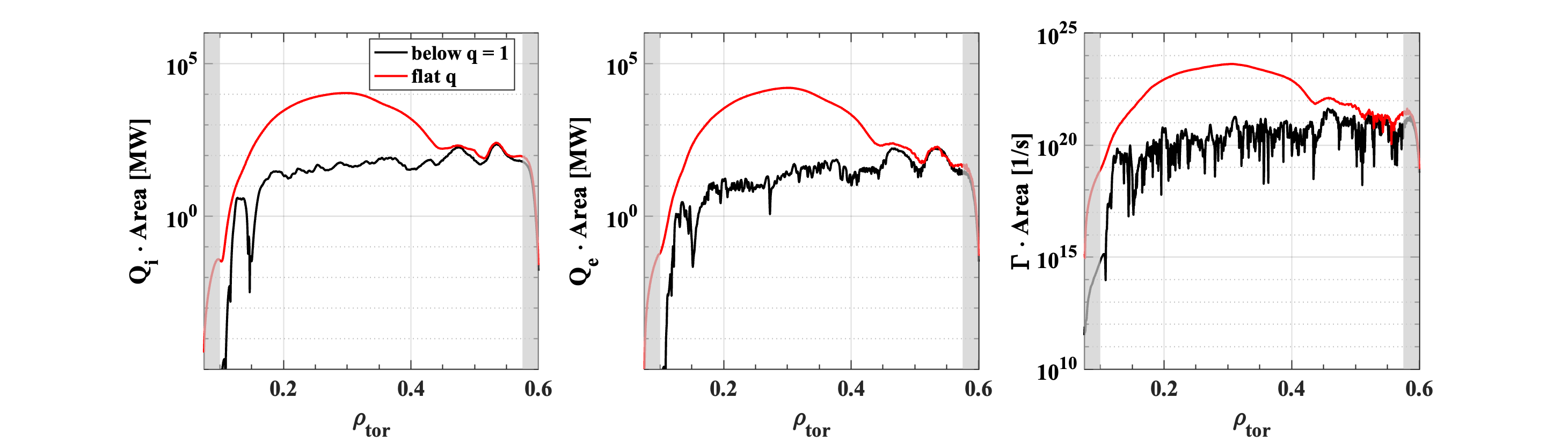}
\par\end{center}
\caption{Comparison of the time-averaged radial profiles of the (a) ion heat flux, (b) electron heat flux (both in MW), and (c) particle flux (in $1/s$) obtained from radially global electromagnetic GENE simulations. The simulations use the steady-state profiles from the global electromagnetic GENE simulation, employing relaxed q (black) and flattened q profile (red). The shaded gray areas denote the buffer regions.}
\label{fig:fig_flat_heat}
\end{figure*}
Fig.~~\ref{fig:fig_flat_heat} clearly shows that the turbulent fluxes increase by two orders of magnitude for the ion heat flux, and by three orders of magnitude for the electron and particle fluxes, leading to a substantial deviation from power balance. This increase is localized within the radial domain $\rho_{tor} < 0.4$, which corresponds to the region where the safety factor profile is flattened, resulting in nearly zero magnetic shear across this entire range. The large increase in turbulent transport is driven by the destabilization of KBMs generating large-scale turbulence structures. These findings indicate that performing flux-matching GENE-Tango simulations with a flat q-profile in the plasma core would yield steady-state profiles significantly lower than those obtained with a safety factor profile below $q = 1$. Such a degradation in core confinement would lead to substantially reduced fusion power. Therefore, identifying strategies to mitigate KBM activity in the post sawtooth crash phases to minimize the fusion power reduction in ITER scenarios is crucial and will be the focus of future investigations, in particular thanks to improved reduced model in integrated modelling frameworks.

\section{What is the relevance of ETG turbulence?} \label{sec10}

Given the relatively large electron temperature gradients obtained in the GENE-Tango steady-state profiles for this ITER baseline scenario, it is important to evaluate the possible contribution of ETG modes to the overall turbulent transport. Up to this point, our analysis has focused solely on ion-scale instabilities, namely ITG, TEM, and MTM/AITG/KBM, thereby neglecting the potential impact of electron-scale turbulence. This raises the question of whether the profile predictions presented in the previous sections may be overly optimistic, as they do not account for a possibly significant flux contribution driven by ETG modes. This investigation is further motivated by previous studies \cite{Maeyama_Nature_2022}, which, under similar plasma conditions, identified a substantial impact of ETGs on electron heat transport. 

As shown in the previous sections, flux-tube (local) simulations can reproduce the turbulent fluxes obtained from more computationally intensive global simulations with reasonable accuracy. Therefore, in this section, we perform both linear and nonlinear flux-tube simulations at the radial location $\rho_{tor} = 0.5$ to evaluate the role of ETGs in the plasma core.

\begin{figure}
\begin{center}
\includegraphics[scale=0.25]{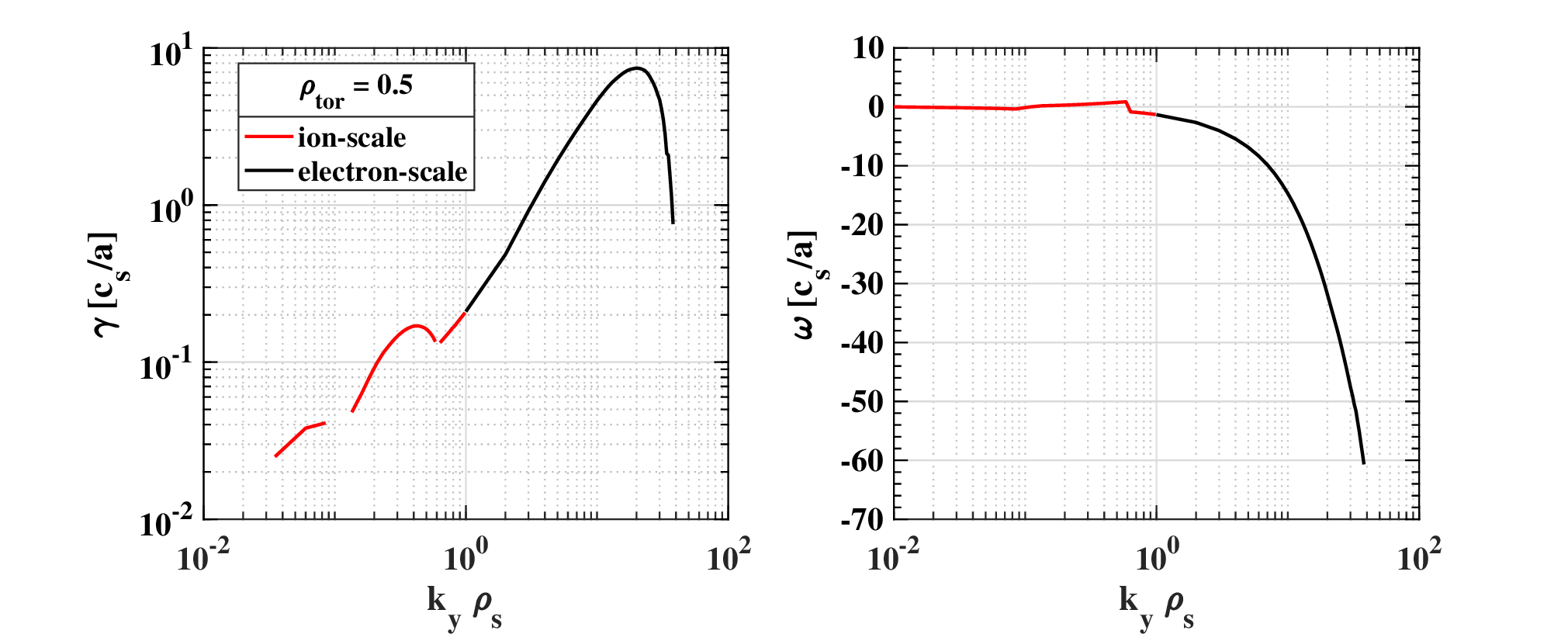}
\par\end{center}
\caption{Comparison of linear growth rates (a) and real frequencies (b) from electromagnetic flux-tube GENE simulations at $\rho_{tor} = 0.5$, covering ion-scales (red) and electron-scales (black). All simulations use thermal profiles extracted from the final state of the global electromagnetic GENE-Tango simulation.}
\label{fig:fig_etg_linear}
\end{figure}

We begin by extending the linear scan presented in Sec.~\ref{sec5} to include electron-scale wave numbers. The results are shown in Fig.~\ref{fig:fig_etg_linear}. 
By evaluating the ratio between the maximum ITG growth rate and its corresponding wave number, and the maximum ETG growth rate and its wave number, we find a value of approximately 0.9, that indicates that the ETG drive is nearly as strong as the ITG one, consistently with the analyses of Ref.~\cite{Maeyama_Nature_2022}. This observation further supports the need for dedicated nonlinear simulations to assess the role of ETG-driven turbulence in the overall electron transport.

For this reason, we perform nonlinear flux-tube (local) GENE simulations resolving the ETG scales. The simulations use the reference resolution and numerical setup introduced in Sec.~\ref{sec2}, with the exception of the bi-normal resolution, which is reduced to $n_{k_y}=36$ with a minimum wave vector of $k_y \rho_e = 0.02$. Note that here the wave vector is normalized to the electron Larmor radius, in contrast to the ion Larmor radius used elsewhere in the manuscript. The radial box size is also increased, reaching 1024 electron Larmor radii. This large domain is necessary to accurately capture the ETG streamers and prevent boundary conditions from influencing the ETG dynamics. Finite-$\beta$ effects and collisions are included in the modelling, using the local Zeff value at the selected flux surface. We perform three sets of simulations: (i) with kinetic ions and the linearized Landau–Boltzmann operator \cite{Crandall_CPC_2020}, (ii) with kinetic ions and the more sophisticated Sugama collision operator \cite{Sugama_PoP_2009}, and (iii) with adiabatic ions, still retaining collisional effects via the linearized Landau–Boltzmann operator and using the same ion-to-electron temperature ratio as in the kinetic cases to check the impact of the collisional operator on the ETG turbulent fluxes.

\begin{figure}
\begin{center}
\includegraphics[scale=0.25]{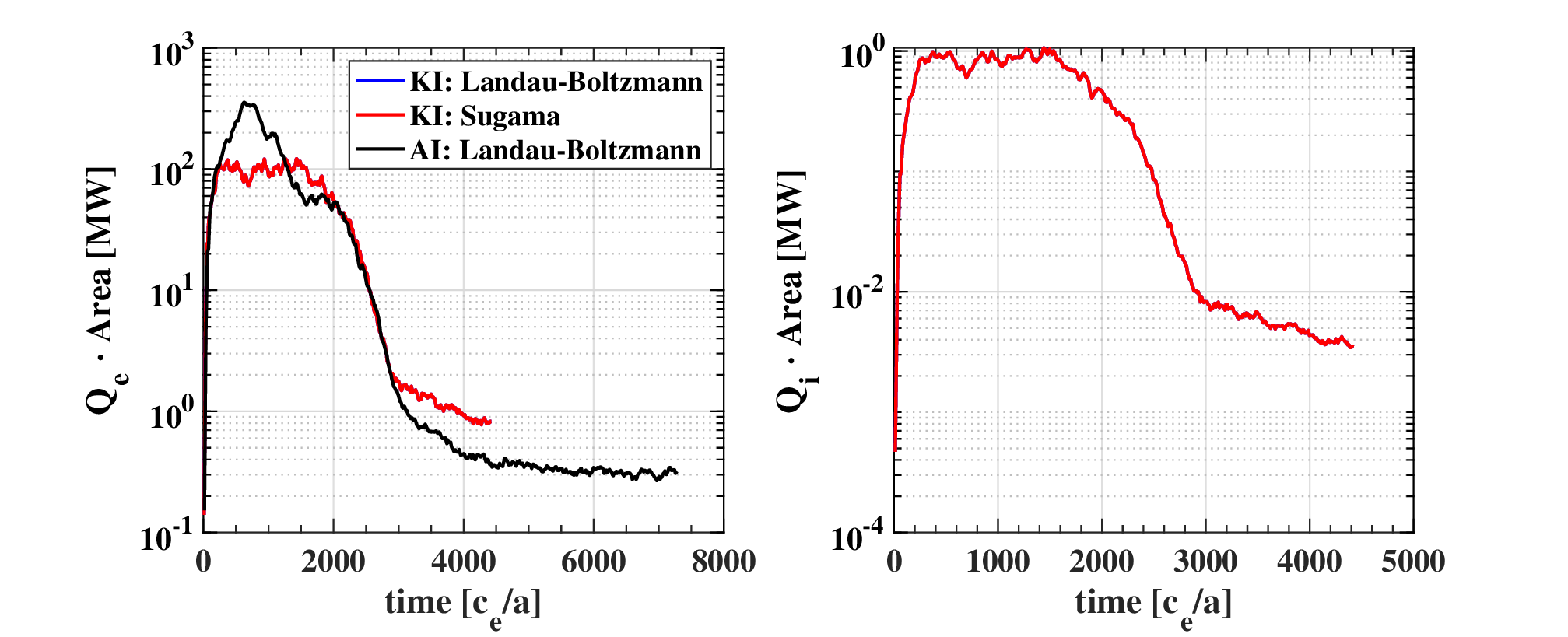}
\par\end{center}
\caption{Time traces of the (a) electron and (b) ion turbulent heat fluxes (in MW) from flux-tube GENE simulations at the radial location $\rho_{tor} = 0.5$, covering only electron scales. The simulations use the steady-state GENE-Tango profiles obtained from the global electromagnetic run. The simulations use kinetic ions with a linearized Landau-Boltzmann collision operator (blue), kinetic ions with the Sugama collision operator (red), and adiabatic ions with the linearized Landau-Boltzmann operator (black).}
\label{fig:fig_etg_flux}
\end{figure}

The results of these simulations are presented in Fig.~\ref{fig:fig_etg_flux}, showing the time traces of both the electron and ion heat fluxes.
Interestingly, Fig.~\ref{fig:fig_etg_flux} shows that, regardless of the simulation setup, the turbulent electron heat flux initially quasi-saturates at very large values ($Q_e \approx 100 MW$), exceeding even those observed in the ion-scale simulations. However, after a long simulated time of approximately $t = 2000 c_e/a$, all simulations undergo a sharp and rapid transition, with the flux dropping by more than two orders of magnitude to values that are negligible compared to those found in the ion-scale cases discussed above. During the initial quasi-stationary phase, turbulence is dominated by radially elongated streamer structures, whereas in the final saturated state it is entirely dominated by zonal flows. 

To better understand these results, we perform a nonlinear scan over the electron collision frequency, ranging from zero up to ten times the nominal value. 
\begin{figure}
\begin{center}
\includegraphics[scale=0.25]{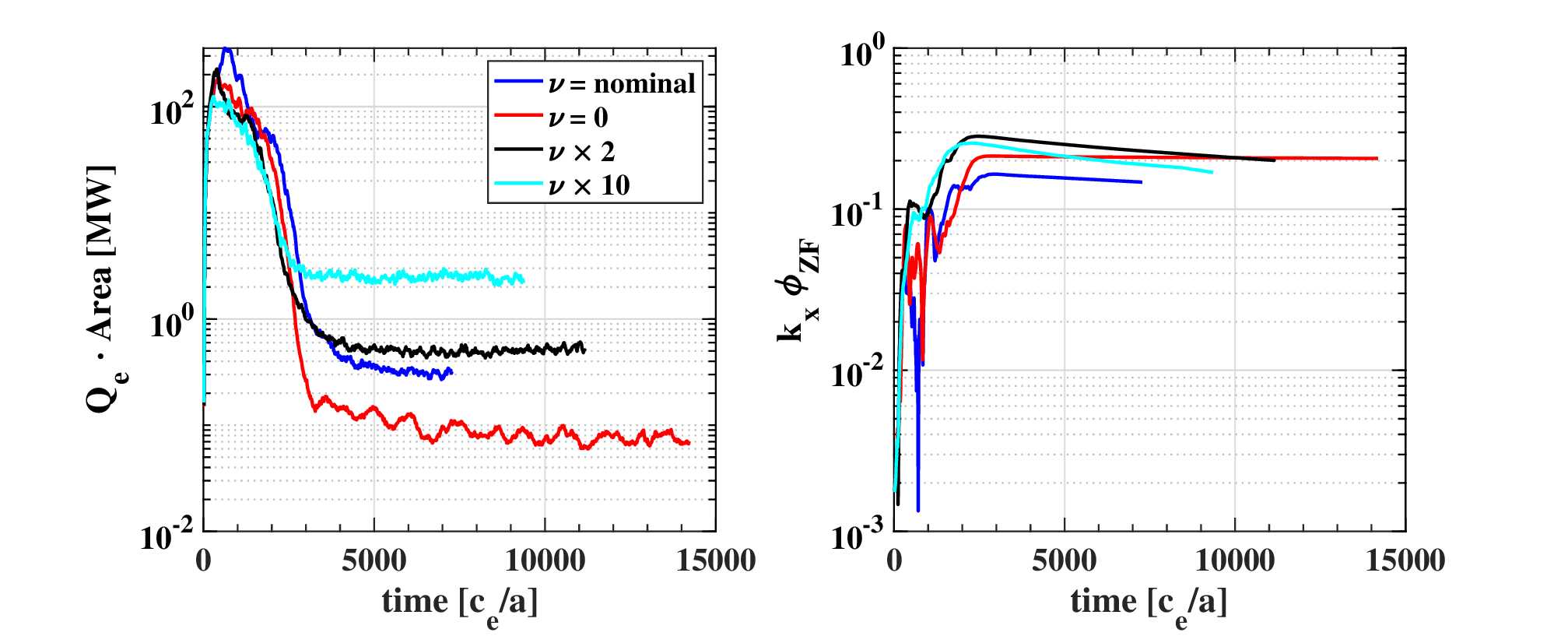}
\par\end{center}
\caption{Time traces of the (a) electron turbulent heat flux (in MW) and (b) zonal component of the radial electric field from flux-tube GENE simulations at the radial location $\rho_{tor} = 0.5$, resolving only electron-scale turbulence at different values of plasma collisionality. All simulations use adiabatic ions and a linearized Landau–Boltzmann collision operator, with background profiles taken from the steady-state global electromagnetic GENE-Tango simulation.}
\label{fig:fig_coll_can}
\end{figure}
For simplicity, these simulations use adiabatic ions and the linearized Landau–Boltzmann collision operator, as Fig.~\ref{fig:fig_etg_flux} shows that the qualitative behavior remains unchanged compared to the more computationally expensive setup with kinetic ions and the Sugama collision operator. The results are presented in Fig.~\ref{fig:fig_coll_can}. This figure clearly shows that, while the initial quasi-stationary phase is largely insensitive to the collisionality level, the final saturated state is strongly affected by it. Specifically, we observe that the electron heat flux in the final state increases almost linearly with the collision frequency, ranging from approximately $Q_e \approx 0.1 MW$ in the collisionless case to $Q_e \approx 2MW$ when the collision frequency is increased by an order of magnitude. As a reference, the expected turbulent fluxes at this radius from ion-scale turbulence are approximately 80 MW for thermal ions and 57 MW for electrons.

These findings are consistent with the results reported by Colyer et al. in Ref.~\cite{Colyer_PPCF_2017}, that investigated ETG turbulence in MAST plasmas. Specifically, in low-collisionality regimes, zonal flows can slowly but steadily grow over time due to the lack of effective collisional damping. Once these flows reach sufficiently large amplitudes, they are able to counteract the turbulence drive. This behavior is illustrated in Fig.~\ref{fig:fig_coll_can}, which shows the temporal evolution of the zonal radial electric field across the different simulations. We observe that the zonal mode amplitude increases gradually and peaks around the time when the electron heat flux begins to drop, eventually settling into a quasi-stationary state. This transition leads to a significantly lower level of saturated electron heat flux. Interestingly, as also found in Ref.~\cite{Colyer_PPCF_2017}, the zonal radial electric field appears largely independent of the collision frequency, even though the heat flux scales almost linearly with it. This can be explained by the fact that, at saturation, the perturbed zonal gradients effectively cancel out the equilibrium background gradients, an effect supported both numerically and analytically in Ref.~\cite{Colyer_PPCF_2017}. Due to the anticipated development of zonal flows, the impacts of ETGs on the global confinement level will be limited, especially for the low collisionality core region, and therefore the observations in the previous sections are likely to be valid while further analyses may be still required to confirm this tentative conclusion

\section{Conclusions} \label{sec11}

In this work, we have presented the first profile prediction of the ITER baseline scenario at 15 MA based on radially global gyrokinetic GENE simulations, leveraging the coupling of the global gyrokinetic code GENE with the transport solver Tango. Our simulations, predict steady-state plasma profiles for $T_e$, $T_i$ and density within $\rho_{tor}=0.6$. The predicted profiles exhibit pronounced density peaking moderated by electromagnetic fluctuations, consistent with the inward particle flux expected in ITG dominated regimes and with empirical scalings from H-mode discharges across multiple devices \cite{Angioni_PPCF2009}.

The predicted fusion gain of $Q=12.2$ aligns well with ITER’s mission goals and previous numerical estimates obtained from local gyrokinetic codes and reduced turbulence models, despite differences in physical assumptions and numerical setup. Our turbulence characterization reveals the critical role of electromagnetic modes, specifically MTMs, EM modes at low wavenumber and AITGs/KBMs at low bi-normal wave numbers in shaping core transport, necessitating high numerical resolution for accurate modeling.

Comparisons with local flux-tube simulations confirm the main qualitative features of global gyrokinetic results but highlight increased stiffness due to the linearization of equilibrium profiles and safety factor in local models, underscoring the importance of global effects in predicting turbulent transport. Moreover, we found that realistic toroidal rotation profiles have negligible influence on turbulent fluxes due to their small magnitude relative to dominant linear growth rates.

Additionally, our results demonstrate that local and global simulations yield similar dominant linear frequencies and mode structures, suggesting that the observed stiffness of the system, make turbulence thresholds critically important. This provides a promising basis for the applicability of reduced models, which is also supported by the good agreement we find with TGLF-SAT2 (using advanced electromagnetic settings \cite{Najlaoui_2025}). Verification of existing reduced models against these global nonlinear results will be further pursued to consolidate these findings. Moreover, the interplay with fusion born alpha particles is currently under investigation and will be addressed in future work.

A key finding of our study is the critical role of the safety factor profile. Flat q profiles with near-zero magnetic shear in the plasma core can destabilize KBMs, thereby enhancing turbulent transport and potentially degrading confinement. Our results therefore suggest that scenarios with a monotonic q profile and stabilized sawteeth are likely to achieve better confinement than flat $q \approx 1$ profiles, where core KBM destabilization can occur. Lastly, while single-scale ETG turbulence initially appears particularly strong, surpassing ion-scale transport levels, it is ultimately quenched on long timescales by zonal flow dynamics under ITER’s expected low collisionality conditions.

\section*{Acknowledgements}

The authors would like to acknowledge insightful discussions with N. Howard, P. Rodriguez-Fernandez, P. Lauber, A. Mishchenko, A. Najlaoui, G. Staebler and F. Zonca. This work has been carried out within the framework of the EUROfusion Consortium, funded by the European Union via the Euratom Research and Training Programme (Grant Agreement No 101052200 — EUROfusion). Views and opinions expressed are however those of the author(s) only and do not necessarily reflect those of the European Union or the European Commission. Neither the European Union nor the European Commission can be held responsible for them. The views and opinions expressed herein do not necessarily reflect those of the ITER organisation. Numerical simulations were performed at the Marconi, Leonardo and JFRS Fusion supercomputers at CINECA, Italy.

\end{document}